\input harvmac 
\input mssymb.tex
\input epsf.tex

\overfullrule=0mm

\newcount\figno
\figno=0
\def\fig#1#2#3{
\par\begingroup\parindent=0pt\leftskip=1cm\rightskip=1cm\parindent=0pt
\baselineskip=11pt
\global\advance\figno by 1
\midinsert
\epsfxsize=#3
\centerline{\epsfbox{#2}}
\vskip 12pt
{\bf Fig. \the\figno:} #1\par
\endinsert\endgroup\par
}
\def\figlabel#1{\xdef#1{\the\figno}}
\def\encadremath#1{\vbox{\hrule\hbox{\vrule\kern8pt\vbox{\kern8pt
\hbox{$\displaystyle #1$}\kern8pt}
\kern8pt\vrule}\hrule}}


\def\IR{\relax{\rm I\kern-.18em R}}
\font\cmss=cmss10 \font\cmsss=cmss10 at 7pt

\font\cmss=cmss10 \font\cmsss=cmss10 at 7pt
\def\IZ{\relax\ifmmode\mathchoice
{\hbox{\cmss Z\kern-.4em Z}}{\hbox{\cmss Z\kern-.4em Z}}
{\lower.9pt\hbox{\cmsss Z\kern-.4em Z}}
{\lower1.2pt\hbox{\cmsss Z\kern-.4em Z}}\else{\cmss Z\kern-.4em Z}\fi}
\def\IN{\relax{\rm I\kern-.18em N}}
\def\b{\circ}
\def\n{\bullet}
\def\gbb{\Gamma_2^{\b \b}}
\def\gnb{\Gamma_2^{\n \b}}

\def\gnn{\Gamma_2^{\n \n}}
\def\gbbbb{\Gamma_4^{\hbox{$\scriptstyle \b \b$}\kern -8.2pt
\raise -4pt \hbox{$\scriptstyle \b \b$}}}
\def\gnnnn{\Gamma_4^{\hbox{$\scriptstyle \n \n$}\kern -8.2pt  
\raise -4pt \hbox{$\scriptstyle \n \n$}}}
\def\gnnnnnn{\Gamma_6^{\hbox{$\scriptstyle \n \n \n$}\kern -12.3pt
\raise -4pt \hbox{$\scriptstyle \n \n \n$}}}
\def\gbbbbbb{\Gamma_6^{\hbox{$\scriptstyle \b \b \b$}\kern -12.3pt
\raise -4pt \hbox{$\scriptstyle \b \b \b$}}}
\def\gbbbbc{\Gamma_{4\, c}^{\hbox{$\scriptstyle \b \b$}\kern -8.2pt
\raise -4pt \hbox{$\scriptstyle \b \b$}}}
\def\gnnnnc{\Gamma_{4\, c}^{\hbox{$\scriptstyle \n \n$}\kern -8.2pt
\raise -4pt \hbox{$\scriptstyle \n \n$}}}
\def\Rbud#1{{\cal R}_{#1}^{-\kern-1.5pt\blacktriangleright}}
\def\Rleaf#1{{\cal R}_{#1}^{-\kern-1.5pt\vartriangleright}}
\def\Rbudb#1{{\cal R}_{#1}^{\circ\kern-1.5pt-\kern-1.5pt\blacktriangleright}}
\def\Rleafb#1{{\cal R}_{#1}^{\circ\kern-1.5pt-\kern-1.5pt\vartriangleright}}
\def\Rbudn#1{{\cal R}_{#1}^{\bullet\kern-1.5pt-\kern-1.5pt\blacktriangleright}}
\def\Rleafn#1{{\cal R}_{#1}^{\bullet\kern-1.5pt-\kern-1.5pt\vartriangleright}}
\def\Wleaf#1{{\cal W}_{#1}^{-\kern-1.5pt\vartriangleright}}
\def\Cleaf{{\cal C}^{-\kern-1.5pt\vartriangleright}}
\def\Cbud{{\cal C}^{-\kern-1.5pt\blacktriangleright}}
\def\Crleaf{{\cal C}^{-\kern-1.5pt\circledR}}


\Title{\vbox{\hsize=3.truecm \hbox{SPhT/02-160}}}
{{\vbox {
\bigskip
\centerline{Combinatorics of }
\centerline{Hard Particles on Planar Graphs}
}}}
\bigskip
\centerline{J. Bouttier\foot{bouttier@spht.saclay.cea.fr}, 
P. Di Francesco\foot{philippe@spht.saclay.cea.fr} and
E. Guitter\foot{guitter@spht.saclay.cea.fr}}
\medskip
\centerline{ \it Service de Physique Th\'eorique, CEA/DSM/SPhT}
\centerline{ \it Unit\'e de recherche associ\'ee au CNRS}
\centerline{ \it CEA/Saclay}
\centerline{ \it 91191 Gif sur Yvette Cedex, France}
\bigskip
\noindent We revisit the problem of hard particles on planar
random tetravalent graphs in view of recent combinatorial 
techniques relating planar diagrams to decorated trees.
We show how to recover the two-matrix model solution to this
problem in this purely combinatorial language. 
\Date{11/02}

\nref\SCH{G. Schaeffer, {\it Bijective census and random 
generation of Eulerian planar maps}, Electronic
Journal of Combinatorics, vol. {\bf 4} (1997) R20; see also
G. Schaeffer, {\it Conjugaison d'arbres
et cartes combinatoires al\'eatoires} PhD Thesis, Universit\'e 
Bordeaux I (1998).}
\nref\BMS{M. Bousquet-M\'elou and G. Schaeffer,
{\it Enumeration of planar constellations}, Adv. in Applied Math.,
{\bf 24} (2000) 337-368.}
\nref\PS{D. Poulhalon and G. Schaeffer, 
{\it A note on bipartite Eulerian planar maps}, preprint (2002),
available at {\sl http://www.loria.fr/$\sim$schaeffe/}}
\nref\PoSc{D. Poulhalon and G. Schaeffer, {\it A bijection for loopless 
triangulations of a polygon with interior points}, proceedings of the 
conference FPSAC'02, Melbourne (2002),
available at {\sl http://www.loria.fr/$\sim$schaeffe/}}
\nref\TUTone{W. Tutte, 
{\it A Census of planar triangulations}
Canad. Jour. of Math. {\bf 14} (1962) 21-38.}
\nref\TUTtwo{W. Tutte, 
{\it A Census of Hamiltonian polygons}
Canad. Jour. of Math. {\bf 14} (1962) 402-417.}
\nref\TUTthree{W. Tutte, 
{\it A Census of slicings}
Canad. Jour. of Math. {\bf 14} (1962) 708-722.}
\nref\TUTfour{W. Tutte, 
{\it A Census of Planar Maps}, Canad. Jour. of Math. 
{\bf 15} (1963) 249-271.}
\nref\CENSUS{J. Bouttier, P. Di Francesco and E. Guitter, {\it Census of planar
maps: from the one-matrix model solution to a combinatorial proof},
to appear in Nucl. Phys. {B} (2002).}
\nref\BIPZ{E. Br\'ezin, C. Itzykson, G. Parisi and J.-B. Zuber, {\it Planar
Diagrams}, Comm. Math. Phys. {\bf 59} (1978) 35-51.}
\nref\DGZ{P. Di Francesco, P. Ginsparg
and J. Zinn--Justin, {\it 2D Gravity and Random Matrices},
Physics Reports {\bf 254} (1995) 1-131.}
\nref\EY{B. Eynard, {\it Random Matrices}, Saclay Lecture Notes (2000),
available at {\sl http://www-spht.cea.fr/lectures\_notes.shtml} }
\nref\HARD{J. Bouttier, P. Di Francesco and E. Guitter, {\it Critical
and tricritical hard objects on bicolourable random lattices: exact solutions}
J. Phys. A: Math. Gen. {\bf 35} (2002) 3821-3854.}
\nref\CONC{M. Bousquet-M\'elou and G. Schaeffer, {\it The degree distribution
in bipartite planar maps: application to the Ising model}, preprint
math.CO/0211070.}

\newsec{Introduction}

Recent progress has been made in the enumeration of various types of planar
maps, using bijective methods relating maps to decorated trees 
[\xref\SCH-\xref\PoSc]. These
techniques are different in nature from the original combinatorial
approach of Tutte [\xref\TUTone-\xref\TUTfour] 
and are much easier to generalize. In particular,
it was shown in Ref. \CENSUS\ how to extend these techniques so as
to recover in a purely combinatorial way the general one-matrix model
solution for the enumeration of planar maps with arbitrary valencies
[\xref\BIPZ-\xref\EY].

After these encouraging results, it is natural to turn to more involved
problems of decorated map enumeration, and try to recover in a purely combinatorial
way the results of more involved matrix models describing statistical
``matter" models defined on random graphs\foot{We use the denomination
``graph" throughout this paper to denote maps, i.e. fatgraphs in
the physics language.} (two-dimensional quantum gravity).    
It is worth mentioning that there is an apparent
obstruction to this type of generalization, as the local interactions on the graphs
become non-local on the corresponding trees.
On the other hand, the known matrix model solutions strongly suggest
by their algebraic form that a simple combinatorial approach to these
problems should exist.

\fig{A sample configuration of hard particles on a planar
tetravalent graph. Empty (resp. occupied) vertices are indicated
by white (resp. black) circles. The particle exclusion rule imposes that no two
occupied vertices are adjacent to the same edge.}{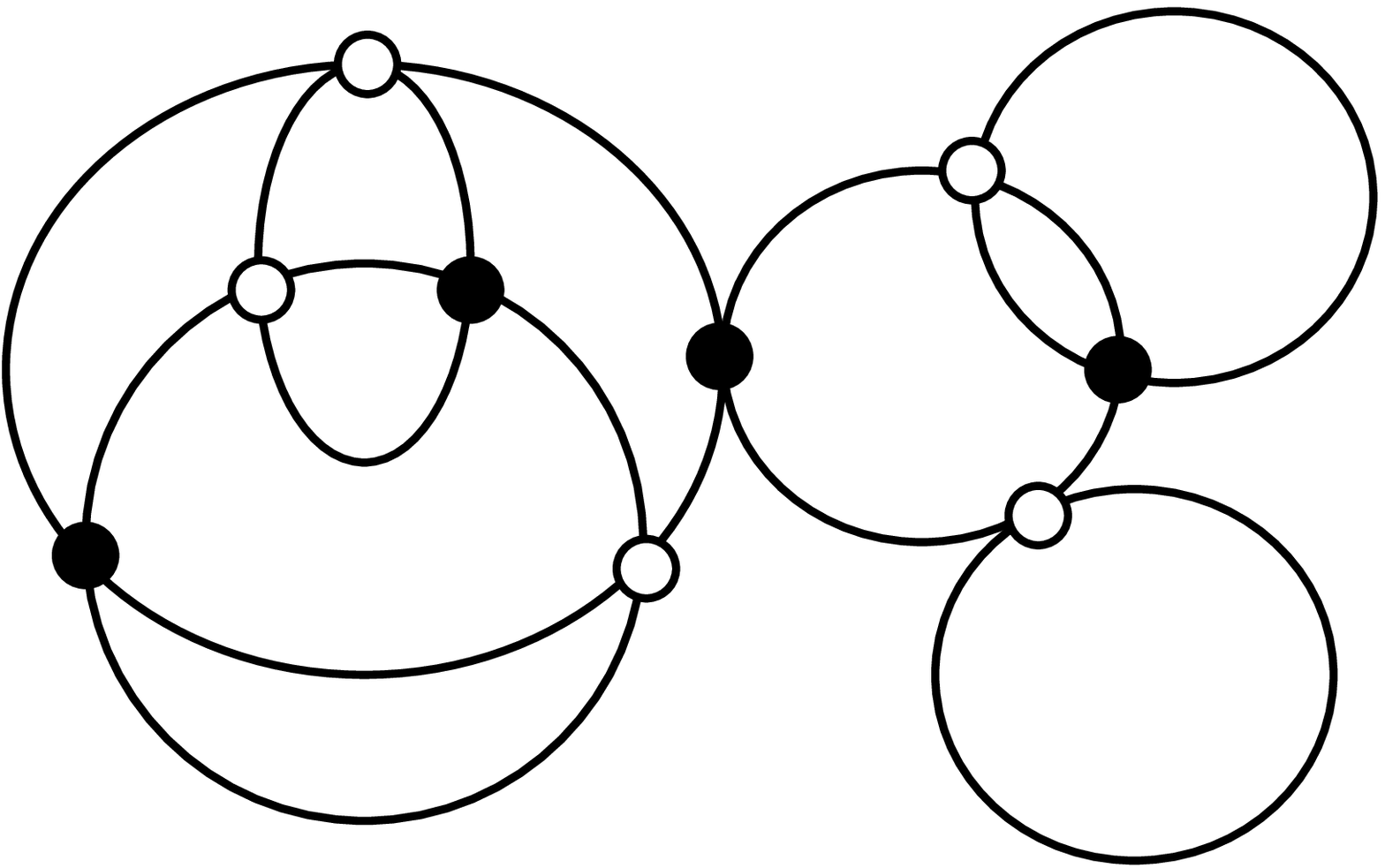}{8.cm}
\figlabel\hardconf

The scope of this paper is to analyze a particular two-matrix model case, namely that describing
{\it hard particles on tetravalent planar graphs}. The configurations of this model
are made of arbitrary tetravalent planar maps with empty or occupied vertices, 
satisfying the hard-particle exclusion rule that no two occupied vertices
may be adjacent to the same edge. Such a configuration is represented in 
Fig. \hardconf\ for illustration. 
This model was solved in Ref. \HARD\ by use of a two-matrix model 
representation. We show here how to recover the solution
of Ref. \HARD\ for the generating function of these
objects {\it in a purely combinatorial manner}, by establishing suitable
bijections between configurations of the model and trees with particles.
The techniques used here are directly borrowed from those of Ref. \CENSUS,
generalizing that of Ref. \SCH.

The paper is organized as follows: In Sect. 2, we recall the
results of Ref. \HARD\ and introduce various types of objects for
which we give closed formulas of the generating functions.
Of particular interest are {\it two-leg} and {\it four-leg} diagrams
as defined below, whose enumeration allows in particular to obtain the 
generating function for all the configurations of hard particles
on tetravalent planar maps.
In Sect. 3, we describe in detail the cutting procedure transforming
two-leg diagrams into decorated trees which we characterize precisely.
The correspondence is proved to be one-to-one by studying the
inverse gluing procedure. We finally use this bijection to enumerate
the two-leg diagrams at hand. Sect. 4 is devoted to the more involved 
case of four-leg diagrams, the enumeration of which is performed through
several steps organized in several subsections. Sect. 5 discusses
an interesting duality property between empty and occupied vertices
of the model. We gather a few concluding remarks in Sect. 6.
The combinatorial counterpart of the more general 
two-matrix models describing bipartite graphs is briefly discussed in Appendix A.

\newsec{Results}

Let us first recall the two-matrix model solution of Ref. \HARD\ to 
the hard-particle model on random planar tetravalent graphs.
The planar free energy $F$, i.e. the generating function for 
configurations of hard particles on connected planar tetravalent (fat)graphs,
was derived and expressed in terms of two functions $R$ and $V$
determined by the following system of equations
\eqn\master{\eqalign{
R&=3 V^2 +9 z R V^2 \cr
V&= \theta + R+3 z R^2+3 z V^3 \cr}}
with the condition that $V=\theta+O(\theta^2)$,
where $\theta$ is the weight per (empty or occupied) vertex and $z$ the
fugacity per particle. 

A more suitable quantity for a combinatorial interpretation
is the generating function for the same configurations  
with a {\it marked oriented  edge}, $E=4 \theta dF/d\theta$, 
where the prefactor $4$ stands for the four possible choices of an edge from a vertex. 
Indeed, as opposed to the case of the free energy, 
the marking of an edge suppresses all symmetry factors
and leaves us with a plain enumeration problem.

\fig{Examples of two- and four-leg diagrams contributing
to $\gnn$ (a), $\gnb$ (b) and $\gnnnn$ (c). Our convention
is to always choose for the external face that containing the
external legs.}{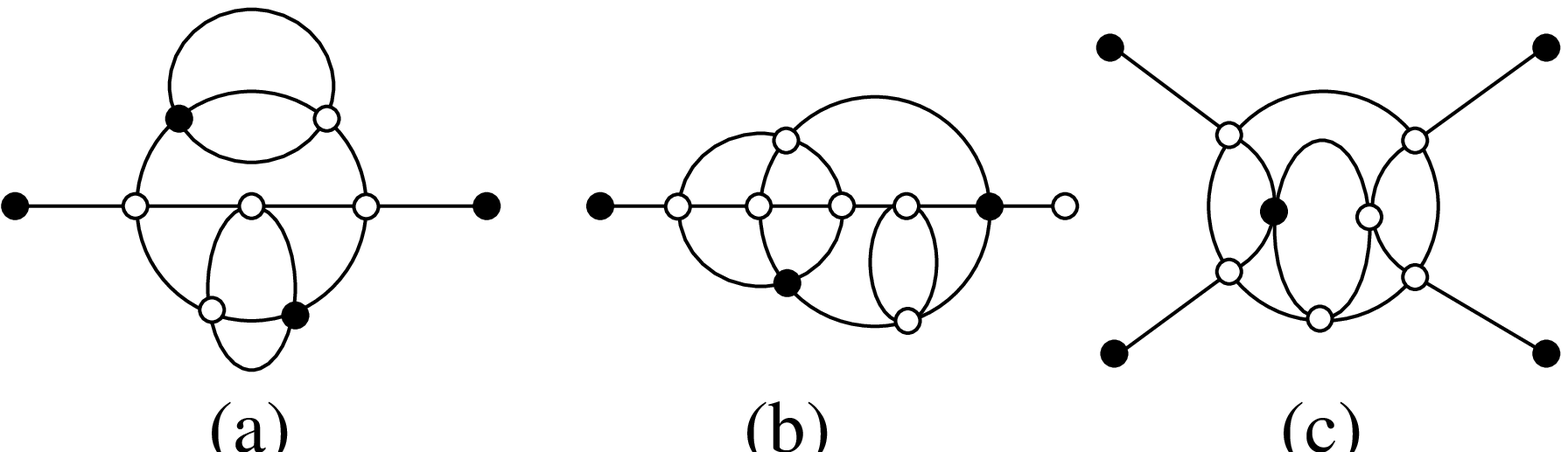}{12.cm}
\figlabel\sampletofor

Moreover, it proves convenient to also consider generating functions for 
diagrams with distinguished (empty or occupied) univalent
vertices (legs) {\it all adjacent to the same (external) face}.
More precisely, let us introduce 
the two- and four-point functions 
$\gnb,\gnn,\gnnnn$
counting respectively two- and four-leg diagrams with the
following external univalent vertices: two occupied ($\n \n $),
one occupied and one empty ($\n \b $) and four 
occupied (\raise -2.pt\hbox{$ \n \n$}\kern -10.pt
\raise 4.5pt \hbox{$ \n \n$}). Examples of such diagrams are depicted
in Fig. \sampletofor. For convenience, we attach
a weight $\sqrt{\theta}$ to each external univalent vertex as 
opposed to the weight $\theta$ per inner vertex. We also decide
{\it not} to attach a factor $z$ to the occupied legs as opposed
to the inner occupied vertices.

\fig{Schematic representation of diagrams contributing
to $E$, $\gnn$ and $\gnb$ according to the empty or occupied
nature of the vertices adjacent to the marked edge or external legs.
Upon cutting the marked edge in the top line into two external legs,
equation (2.2) is obtained.}{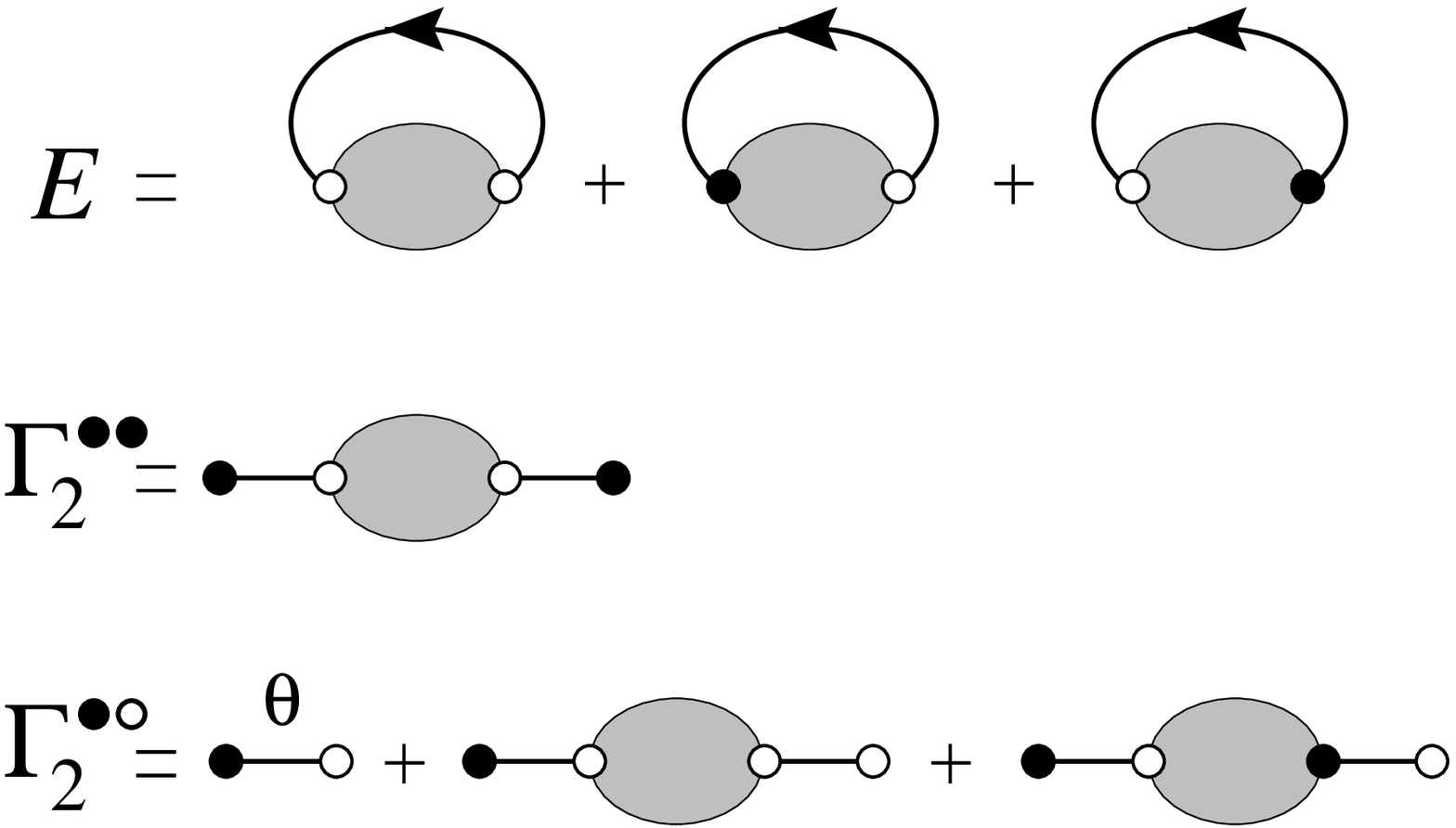}{8.cm}
\figlabel\expressE
\fig{Schematic representation of diagrams contributing to $\gnn$, $\gnnnn$
and $\gnb$ according to the empty or occupied nature
of the vertices adjacent to the external legs. Upon erasing the
occupied vertex on the right of the second line, equation (2.3) is
obtained.}{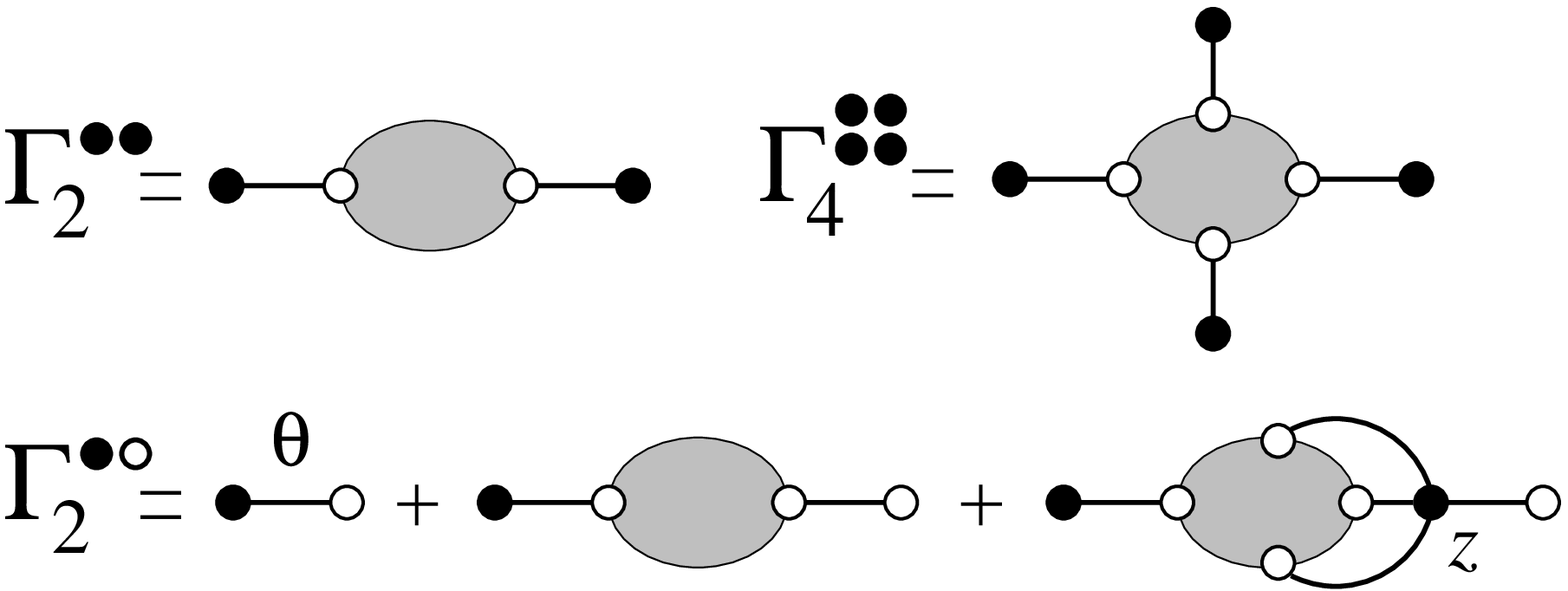}{9.cm}
\figlabel\expressGa

The generating function $E$ is
expressed through   
\eqn\energ{ E= {2 \gnb -\gnn -2 \theta \over \theta} } 
by simply considering all possible configurations of the two
vertices adjacent to the marked oriented edge (see Fig. \expressE)
The two- and four-leg diagrams are further related via 
\eqn\twofour{ \gnb = \theta+\gnn+ z \gnnnn }
according to the configuration of the vertex connected to
the empty external vertex in a diagram of $\gnb$ (see Fig. \expressGa). 

The aim of this paper is to provide a combinatorial interpretation
for the functions $R$ and $V$ as generating functions for decorated trees
and to derive in a purely combinatorial manner the following
expressions for two- and four-point functions 
\eqn\gamtwo{ \gnn = R- {V^3+z R^3 +6 z R V^3\over \theta} }
\eqn\gamfour{ \gnnnn= V^3 +2 R^2 -3{ z V^6+ R V^3+z R^4+7 R^2 V^3\over \theta} }
which immediately lead to compact expressions for $\gnb$ via eqn. \twofour\
and finally $E$ via eqn. \energ.
These relations may alternatively be derived within the framework of the
two-matrix model of Ref. \HARD\ but without any clear combinatorial interpretation
yet.

\newsec{Two-leg diagrams}

In this section we address the simplest case of 
two-leg diagrams with both legs occupied, generated by $\gnn$.
As in Ref. \CENSUS, we will construct a correspondence between
these diagrams and suitably decorated trees whose enumeration
leads to a combinatorial interpretation of $R$ and $V$
as defined through eqn. \master, and to the formula \gamtwo\ for $\gnn$. 

\subsec{Cutting procedure: from diagrams to trees}

\leftline{\sl 3.1.1. Iterative algorithm}
\nobreak
\fig{The cutting of a two-leg diagram contributing to $\gnn$ into
a tree according to the rules explicited in the text. 
The edges of the diagram are visited in counterclockwise
direction starting from the incoming external leg (top vertex).
After one turn, those edges marked in (a) have been cut into
bud-leaf pairs. Buds (resp. leaves) are represented by black 
(resp. white) arrows. The procedure is repeated a second (b) and third (c)
time until no edge may be cut anymore. The desired tree is obtained 
(d) by replacing the external legs by leaves.}{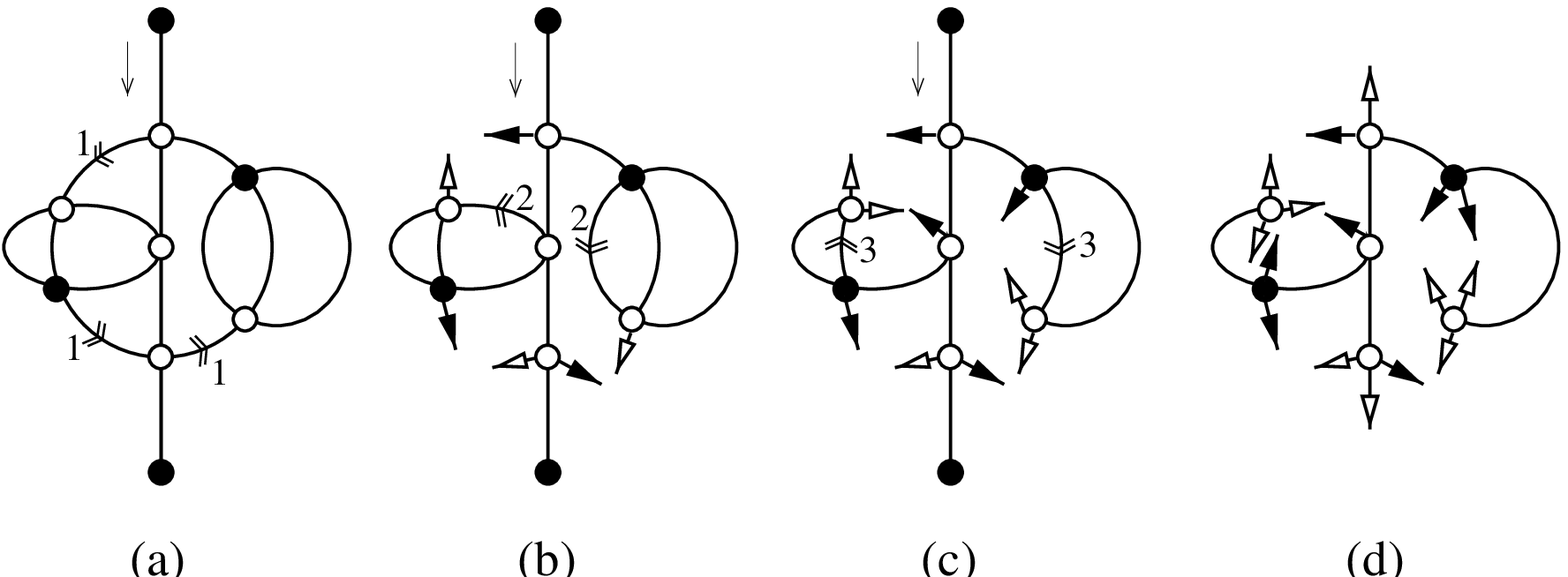}{13.5cm}
\figlabel\cutme

Starting from a two-leg diagram with both legs occupied,  
distinguished as in- and out-coming ones, we apply the following 
edge-cutting algorithm. Starting from the incoming leg, we 
successively visit all edges and vertices adjacent to the external face in
{\it counterclockwise} direction. At each step, the visited edge is
cut iff 
\item{(i)} the remaining diagram is not disconnected and 
\item{(ii)} the next visited vertex is {\it empty}.
\par
\noindent Each cut edge is split into a pair of half-edges to which we attach
respectively a black and a white label (represented by arrows in Fig. \cutme), 
and that we refer to from now
on as a {\it bud} (half-edge with a black label) and a {\it leaf}
(half-edge with a white label). Note that the bud is attached to the vertex
visited prior to the edge, while the leaf is attached to the next one.
This has the effect of merging a number of faces with the external one. 
The procedure is then iterated on the new (cut) diagram and stops when
all faces have been merged. Indeed, the particle exclusion rule
ensures that no unmerged face can remain: if such faces existed, 
among the edges separating the 
external face from unmerged ones, at least one would point to  
an empty vertex in counterclockwise direction, 
clearly contradicting the cutting process.
Finally, we replace the two univalent vertices by leaves.
An explicit example of this procedure is depicted in Fig. \cutme.

\leftline{\sl 3.1.2. Equivalent cutting procedure via leftmost minimal paths}
\nobreak
\fig{Equivalent cutting procedure of the diagram of Fig. \cutme, 
by use of leftmost minimal paths. These are oriented paths of
minimal length connecting the external face to all the other faces.  
When crossing an edge, the vertex on the right has to be empty.
When several minimal paths to a given face exist, we choose the leftmost 
one with respect to the incoming leg.
}{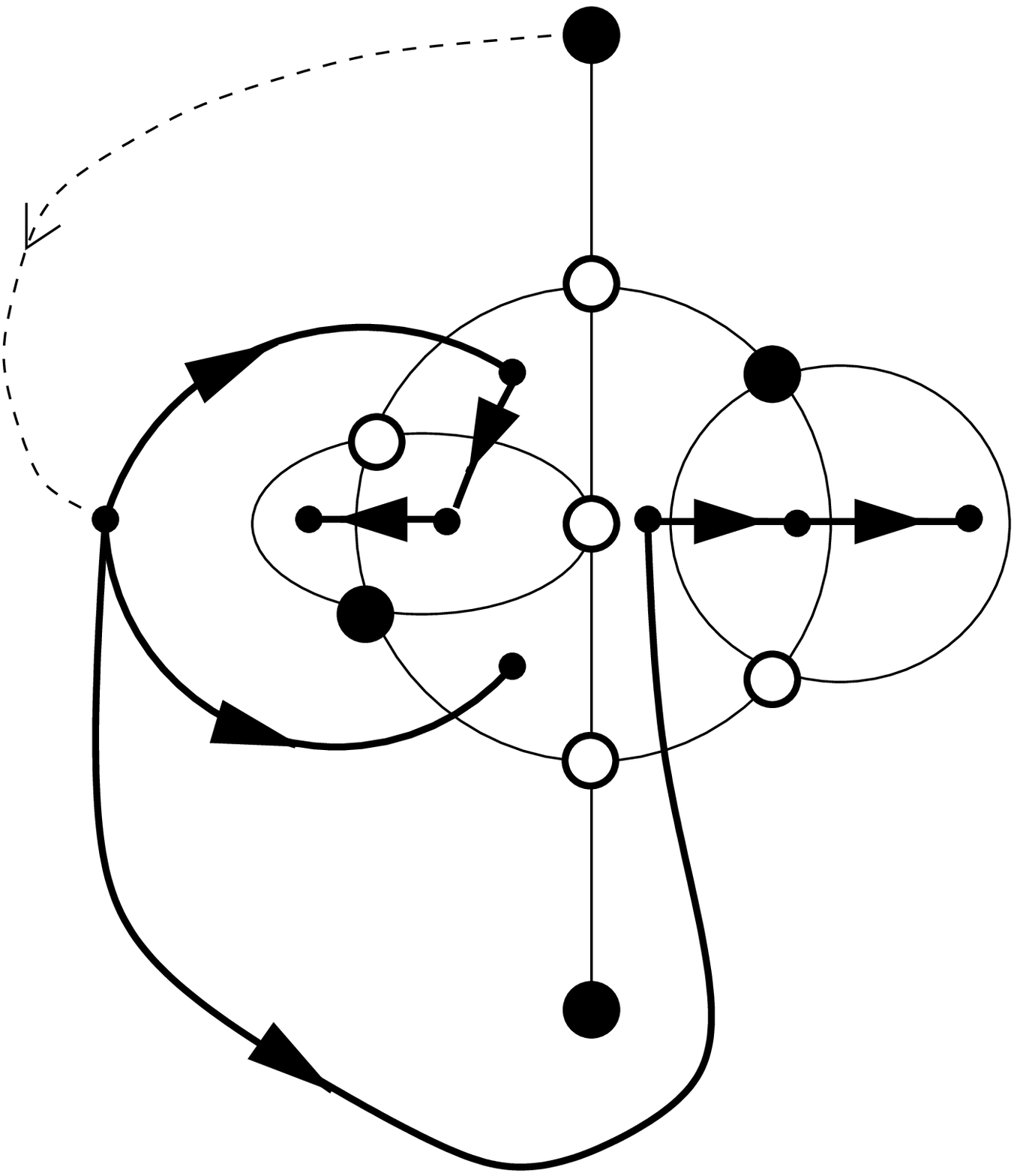}{6.cm}
\figlabel\minimal

Alternatively, the above cutting algorithm may be expressed 
in a dual language by introducing the notion of {\it leftmost 
minimal paths} defined as follows.   
A {\it path} $\{e\}$ between two faces $f$ and $f'$ is a sequence of edges 
$e_1,e_2,...,e_k$, where $e_i$ is adjacent to two faces $f_{i-1}$
and $f_i$, with  $f_0=f$, $f_k=f'$ (it is also clearly a path
drawn on the dual diagram between the corresponding dual vertices).
We further impose the orientation constraint that, when going from $f_{i-1}$
to $f_i$ across $e_i$, {\it the vertex on the right of $e_i$ is empty}. 
Such a path between any two given faces always exists as,
by virtue of the exclusion rule,
any edge with an occupied vertex to its right may be
bypassed to the right, hence crossing only edges with an
empty vertex to their right. 
A {\it minimal path} between two given faces $f$ and $f'$ is such a path 
with minimal length $k$. 
Let us now fix the external face $f_0$ as origin.  Two minimal
paths $\{e\}$ and $\{e'\}$  from $f_0$ to any given face $f$ may be 
compared by examining the mutual position (left or right) 
of their first differing edge    
$e_i\neq e_i'$ with respect to the previous one $e_{i-1}=e_{i-1}'$
(with the convention that $e_0$ is the incoming edge): 
$\{e\}$ is said to lie on the left (resp. right) of $\{e'\}$ if
$(e_{i-1},e_i,e_i')$ appear in clockwise (resp. counterclockwise) 
order around $f_{i-1}$ (with the opposite convention for the external
face $f_0$ due to the planar representation with a point at infinity). 
This total order on minimal paths from $f_0$ to $f$  
allows to define the {\it leftmost minimal path} from $f_0$ to $f$. 
It is easily seen that all the edges belonging to leftmost minimal
paths are those to be cut in the above cutting algorithm,
as illustrated in Fig.\minimal. 
Indeed, any edge $e_i$ of a leftmost minimal path $\{e\}$ is cut 
during the $i$-th step of the cutting algorithm, thus merging the
face $f_i$ with the external one (among the edges separating
$f_i$ from $f_{i-1}$, $e_i$ is the first edge encountered 
in counterclockwise order at step $i$). Note that the leaf
is attached to the (empty) vertex on the right of $e_i$ 
when going from $f_{i-1}$ to $f_i$. 
Conversely, given in the two-leg diagram a face $f_i$
merged at step $i$ by cutting an edge $e_i$,
the leftmost minimal path from $f_0$ to $f_i$ is inductively
given by appending $e_i$ to the leftmost minimal path from $f_0$ to 
$f_{i-1}$ (the face separated from $f_i$ by $e_i$).

\subsec{Characterization of the resulting trees}


The aim of this section is to show that the trees resulting from the cutting procedure
of previous section satisfy the following properties:
\item{(UR1)} These trees are made of tetravalent empty or occupied regular vertices, inner edges
satisfying the hard-particle exclusion rule, and two kinds of endpoints (univalent
vertices): leaves and buds
\item{(UR2)} No leaf is connected to an occupied vertex
\item{(UR3)} Attaching a {\it charge} $+1$ to leaves and $-1$ to buds, the total charge
of the trees is $+2$
\item{(UR4)} Cutting any inner edge separates the trees into two pieces
such that a piece starting with an empty vertex has a non-negative total charge
\par
\noindent Such trees will be called {\it unrooted} R-{\it trees}.

Properties (UR1) (UR2) follow clearly from the iterative
cutting algorithm. 
Attaching a charge $+1$ to leaves and $-1$ to buds, it is also clear that
the total charge is $+2$ (UR3), as leaves and buds appear in pairs, except for the two
leaves replacing the former legs. 
We are left with the task of proving (UR4). As an illustration, the reader may 
check (UR4) on the tree of Fig. \cutme\ (d).

\fig{An example of configuration of buds and leaves around
an inner edge $e$ in the tree resulting from the cutting procedure.
The edge $e$ separates the tree into two parts $T_1$ and $T_2$, with $T_1$
containing the former incoming leg (IN). In
the original diagram, $e$ separates two faces $f$ and $f'$. The leftmost 
minimal paths from the external face $f_0$ to these faces are
$\{e\}=(e_1,e_2,e_3)$ and $\{e'\}=(e_1',e_2',e_3',e_4')$ respectively 
(with the former lying on the left of the latter), which are cut edges 
during the procedure. In
the case depicted here, the former outcoming leg (OUT)
belongs to $T_1$, which is necessarily the case when $\{e\}$
and $\{e'\}$ have some common edges (here $e_1=e_1'$).}
{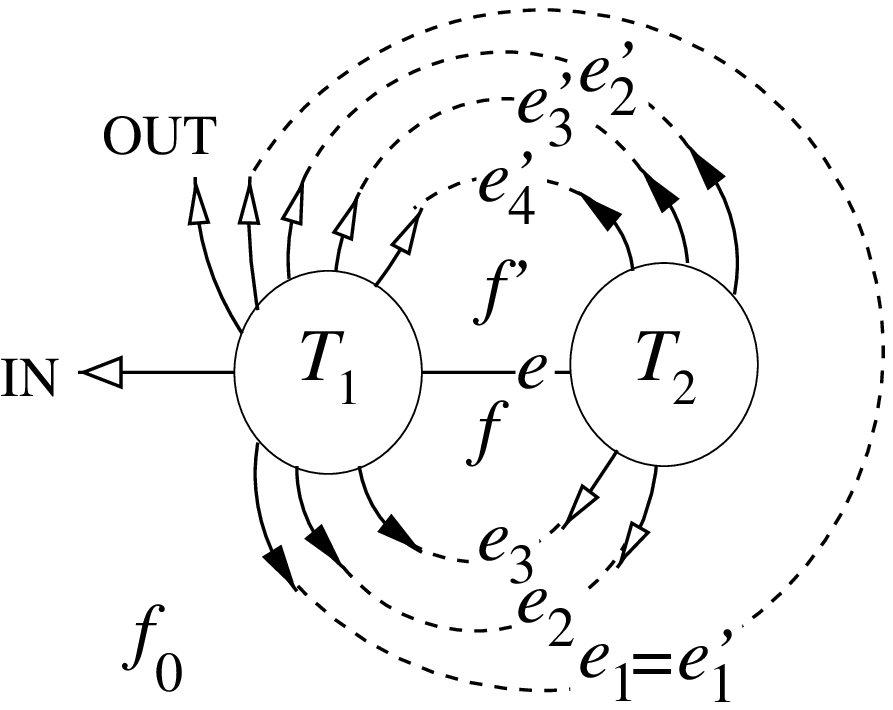}{6.cm}
\figlabel\generic

Let us consider an inner edge $e$ of a tree obtained by cutting 
a two-leg diagram $\Gamma$ as above. This edge separates
the tree into two parts $T_1$, $T_2$ where say $T_1$ contains
the former incoming edge.  
We denote by $f$ and $f'$ the two faces
adjacent to $e$, and by $e_1,e_2,...,e_k$ and $e_1',e_2',...,e_{k'}'$
their respective leftmost minimal paths from the external face $f_0$.
Without loss of generality, we assume that $\{e\}$ is on the left of
$\{e'\}$, as defined above by comparing the position of their first 
differing edges $e_\ell \neq e_\ell'$ ($e_i=e_i'$, $i=0,1,2,...,\ell-1$).
As apparent from Fig. \generic, 
this amounts to assuming that when going from $f$ to $f'$ across $e$,
$T_1$ is on the left.

Starting again from the original two-leg diagram $\Gamma$, we may
equivalently first cut
the edges $e_\ell,e_{\ell+1},...,e_k$ and $e_\ell',e_{\ell+1}',...,e_{k'}'$.
The edge $e$ now clearly separates the diagram into two pieces ${\tilde T}_1$
and ${\tilde T}_2$, which after completing the cutting algorithm turn into
$T_1$ and $T_2$ respectively. The completion of the algorithm
consists in only cutting edges 
within  ${\tilde T}_1$ or ${\tilde T}_2$ (in particular, the cutting of the common edges
$e_1,e_2,...,e_{\ell-1}$ only affects ${\tilde T}_1$), hence the corresponding bud-leaf
pairs do not contribute to the total charge of either piece, i.e. 
$q(T_i)=q({\tilde T}_i)$, $i=1,2$. 
As the edges $e_\ell,e_{\ell+1},...,e_k$ are replaced by buds in ${\tilde T}_1$
and leaves in ${\tilde T}_2$, while $e_\ell',e_{\ell+1}',...,e_{k'}'$
are replaced by buds in ${\tilde T}_2$ and leaves in ${\tilde T}_1$, the
total charges read respectively $q(T_1)=1+\epsilon+(k'-\ell+1)-(k-\ell+1)=1+\epsilon+k'-k$
and $q(T_2)=(1-\epsilon)+(k-\ell+1)-(k'-\ell+1)=1-\epsilon+k-k'$, where 
$\epsilon=1$ if the former outcoming leg lies in $T_1$
and $\epsilon=0$ if it lies in $T_2$.

\fig{A schematic proof of the characterization (UR4) stating 
that any piece of tree starting with a white vertex has a non-negative charge.
The edge $e$ separates the tree into two parts $T_1$ and $T_2$ where
$T_1$ contains the former incoming leg. Among $T_1$ and $T_2$, we can pick 
one starting with an empty vertex. The different cases correspond to
the positions of both the former outcoming leg and the selected empty
vertex. The charges of $T_1$ and $T_2$ depend on $k$ and $k'$, namely the
lengths of the minimal paths from the external face to the faces adjacent 
to $e$, as explained in the text. The condition of leftmost minimal paths
yields constraints on the respective values of $k$ and $k'$ as indicated in
the figure, which in turn translate into the desired constraint 
on the charge of the selected piece.}{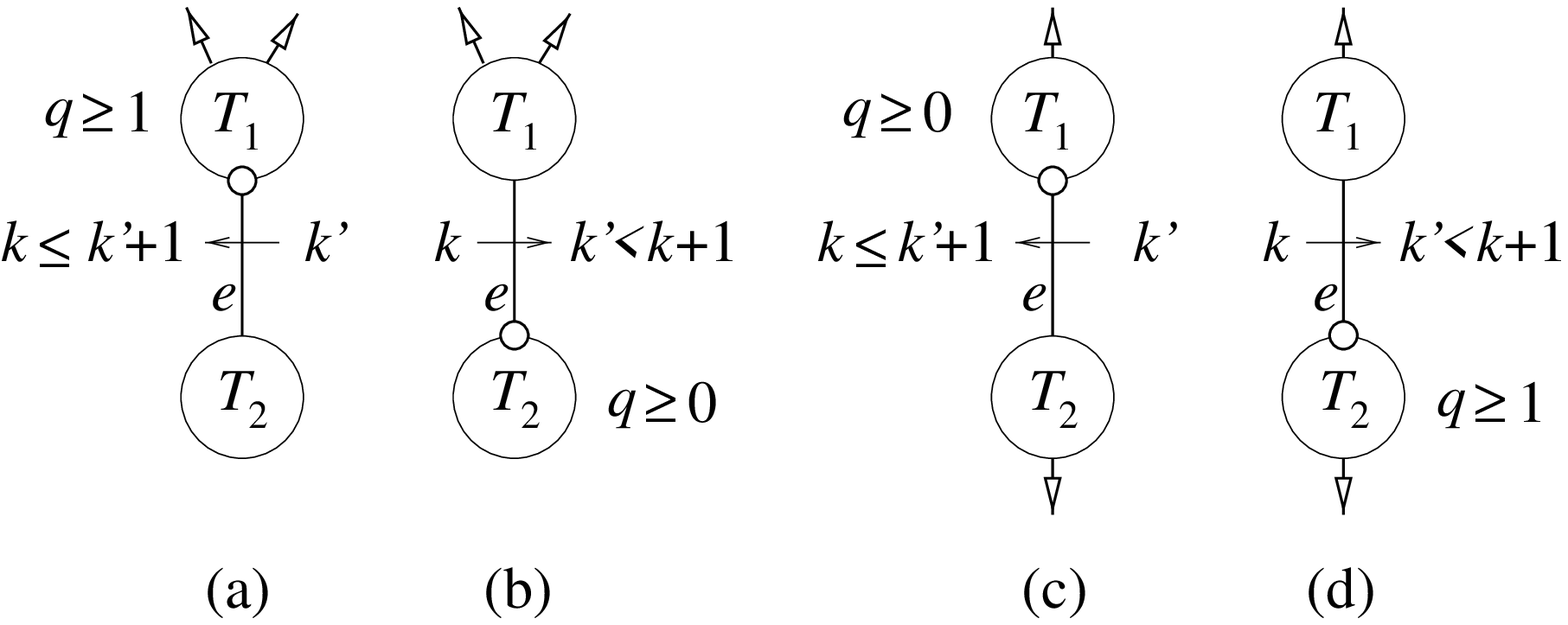}{12.cm}
\figlabel\newrule

We now turn to the inspection of the various possible environments around
the edge $e$. 
The particle exclusion rule implies that at least one part $T_1$ or $T_2$
starts with an empty vertex. Picking such an empty vertex, we are left
with the four possible cases depicted in Fig. \newrule,
where the first two cases (a--b)
correspond to $\epsilon=1$ while the last two (c--d) correspond to $\epsilon=0$.
For each case,
we have also indicated with an arrow a possible way of crossing $e$
which implies restrictions on the values of $k$ and $k'$ as $e$ does not belong 
to a leftmost minimal path. 
In cases (a) and (c), we have $k\leq k'+1$, otherwise the minimality condition 
would be violated by allowing for a shorter path to $f$ from $f_0$ passing
through $f'$ and crossing $e$. In cases (b) and (d), we have $k'\leq k+1$
by a similar minimality argument, but moreover $k'=k+1$ is forbidden
by the leftmost condition. All these restrictions turn into lower bounds for the 
charge $q$ of the piece attached to the selected empty vertex. In all cases,
we find that $q$ is necessarily non-negative, from which (UR4) follows.

\fig{Proof of the equivalence of (UR4) and (UR4') for tetravalent trees, 
as explained in the text.
If $m\geq 2$, one of the $q$'s has to be $\geq 3$ and the complementary part
has negative charge and starts with an empty vertex, in contradiction
with (UR4).}{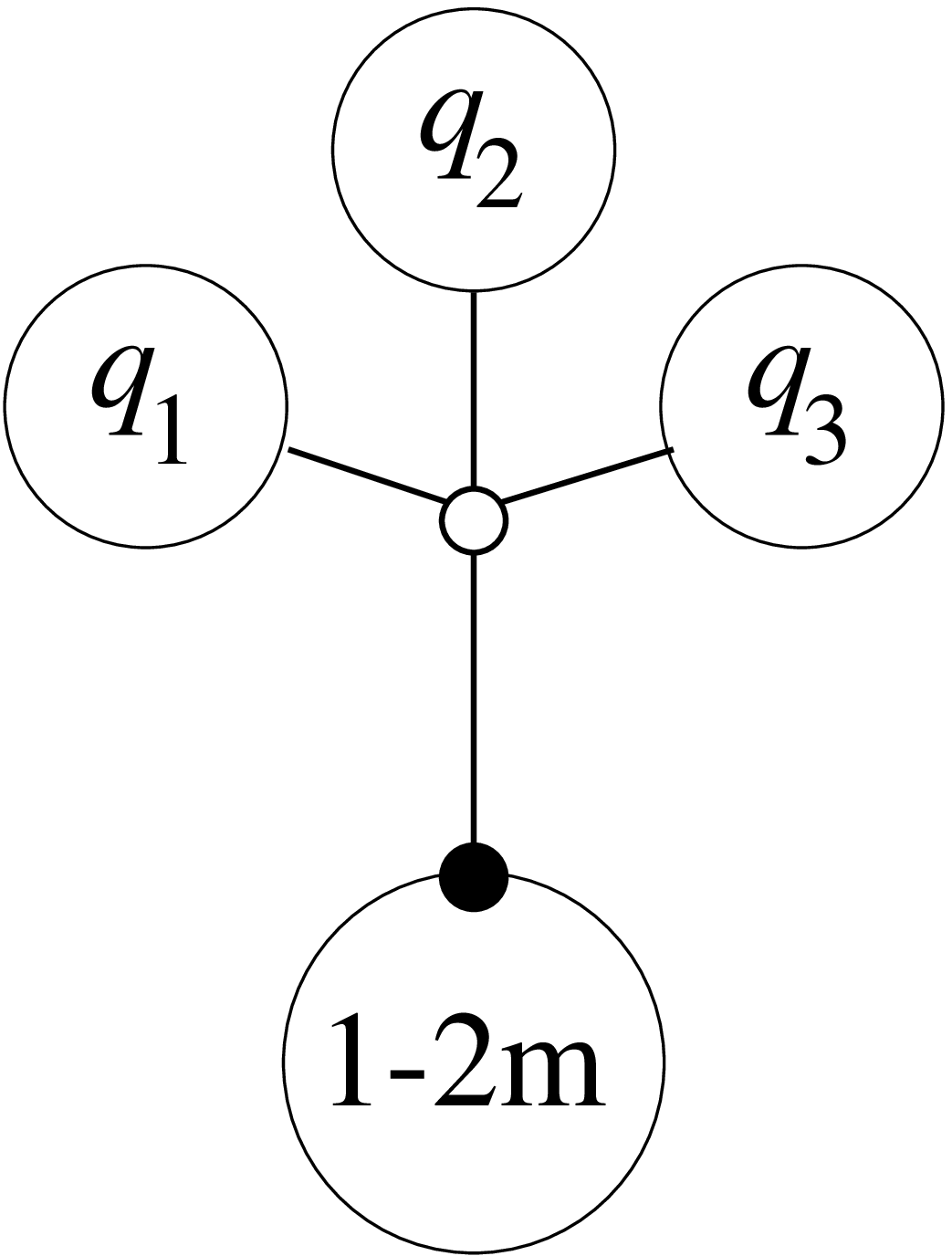}{4.cm}
\figlabel\negq
 
To conclude this section, note that the condition (UR4) above may be 
equivalently replaced by the apparently stronger condition
\item{(UR4')} Cutting any inner edge separates the trees into two pieces of
charges $(+1,+1)$ or $(-1,+3)$, the piece of charge $-1$ always starting
with an occupied vertex
\par  
\noindent To see this equivalence, we first remark that 
since the pieces at hand are trees with only tetravalent
inner vertices, viewing buds, leaves and the half-edge obtained by cutting
the inner edge as external legs, their total number of legs $\#$(buds)$+\#$(leaves)$+1$
is necessarily even, hence their total charge $\#$(leaves)$-\#$(buds) is odd.
As the total charge of the unrooted R-tree is $+2$, the only possible cases are $(+1,+1)$
or $(-(2m-1),2m+1)$, $m=1,2,...$, the piece of charge $-(2m-1)$ necessarily
starting with an occupied vertex, while that of charge $2m+1$ starts
with an empty one. Let us now show that only $m=1$ is allowed.
By contradiction, assume $m\geq 2$. As shown in Fig. \negq, cutting the empty 
vertex leaves us with four
pieces of odd charges $(1-2m),q_1,q_2,q_3$, such that $q_1+q_2+q_3=2m+1\geq 5$. This
implies that at least one of the $q$'s say $q_i\geq 3$, hence the complementary piece
both has negative charge $2-q_i\leq -1$ and starts with an empty vertex, in
contradiction with (UR4).

\subsec{Inverse gluing procedure: from unrooted R-trees to two-leg diagrams}

We have proved in the previous section that cutting a two-leg
diagram with both legs occupied yields an unrooted R-tree.
Let us now show that the correspondence is one-to-one
between two-leg
diagrams with two distinguished occupied legs and unrooted R-trees
{\it with a marked leaf at depth $0$} as defined below. 

\leftline{\sl 3.3.1. Gluing procedure}
\nobreak
\fig{Inverse procedure: gluing back the tree of Fig. \cutme (d)
brings back the original two-leg diagram of Fig \cutme (a) by
identifying the unmatched (depth 0) leaves with the external legs. One
of these must be chosen as the incoming leg.}{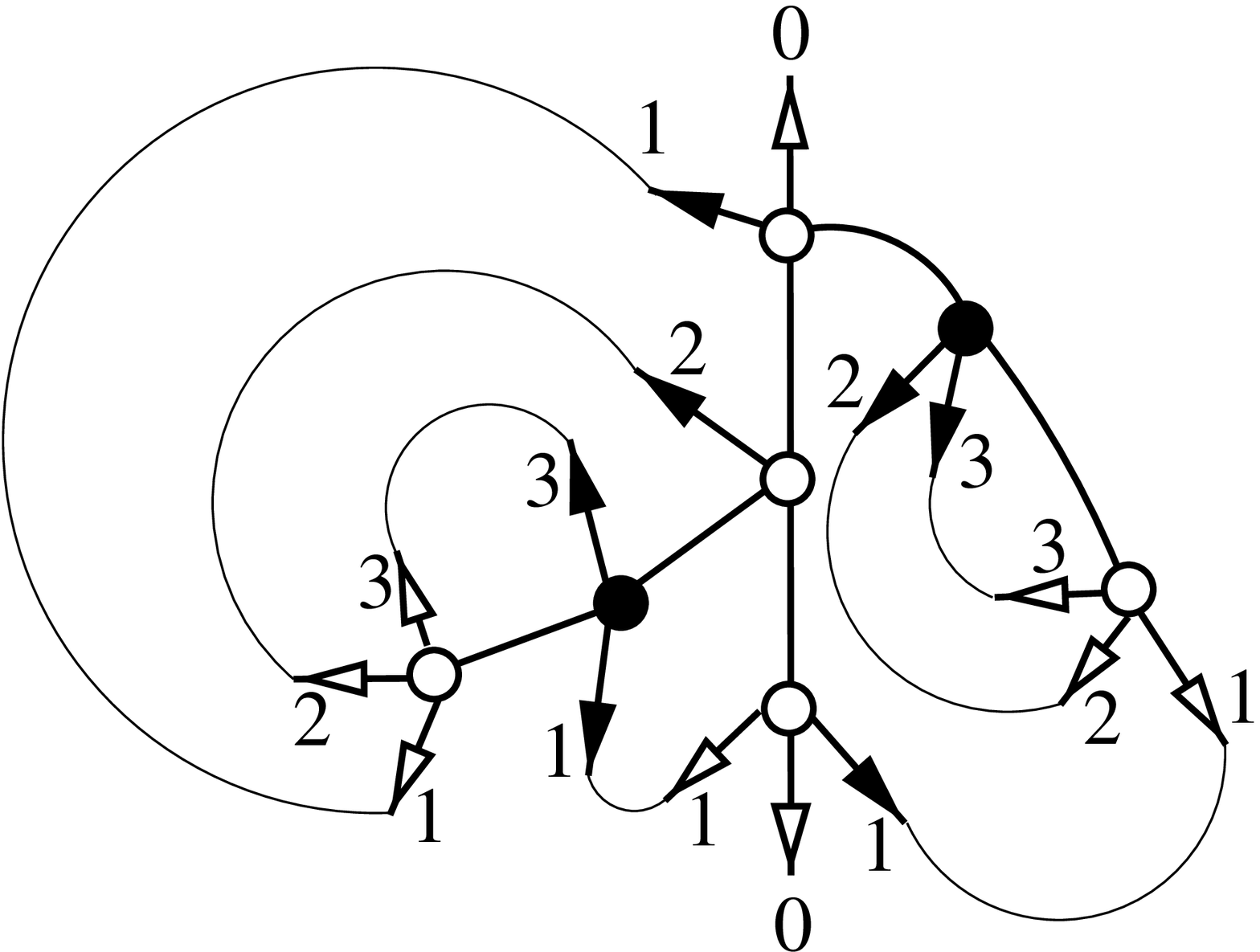}{7.cm}
\figlabel\gluing

Starting from an unrooted R-tree, let us iteratively connect each bud
to the nearest available leaf in counterclockwise direction around the tree
so as to form edges in such a way that the resulting graph is planar
(see Fig. \gluing\ for an example).
This procedure is clearly unique and leaves us with exactly two
unmatched leaves, which are given the depth $0$. The edges
obtained by bud-leaf connections form two systems of arches separated
by the two unmatched leaves, each bud and leaf receives the depth of
the corresponding arch, starting with depth $1$ for external arches. 
Replacing the marked unmatched leaf with the incoming occupied leg and the other
unmatched one with the outcoming occupied leg, we end up with a planar two-leg diagram
with empty and occupied vertices 
which clearly satisfy the hard-particle exclusion rule. Indeed, the rule is obviously
satisfied along inner edges of the original tree, while for the new edges it follows
from the property (UR2) above that leaves are only connected to empty vertices. 

\leftline{\sl 3.3.2. Proof that gluing is the inverse of cutting}
\nobreak
It remains to be shown that the gluing procedure is the inverse of the cutting
algorithm of the previous section. 
First let us note that cutting a two-leg diagram into a tree and connecting
the tree's bud-leaf pairs according to the gluing prescription clearly brings
back the original diagram. We are left with the task of proving that conversely
gluing an unrooted R-tree into a two-leg diagram and applying the cutting algorithm
brings back the original unrooted R-tree.
This boils down to showing that the bud-leaf pairs of any unrooted 
R-tree form the edges of the leftmost minimal paths
from the external face to the inner faces of the diagram obtained by connecting them. 

\fig{Justification of the inverse gluing procedure. As explained in the
text, we assume by contradiction that some inner edge $e_k$ in an unrooted
R-tree belongs to a leftmost minimal path $\{e\}$ in the diagram constructed 
by the gluing procedure. Assuming that the value of $k$ is minimal, we are
in either situation (a) or (b) depending on the position of the former incoming
leg. As explained in the text, the leftmost minimal condition implies bounds
on the charges of $T_1$ and $T_2$ which, because of (UR4), are in contradiction
with the orientation rule for minimal paths.}{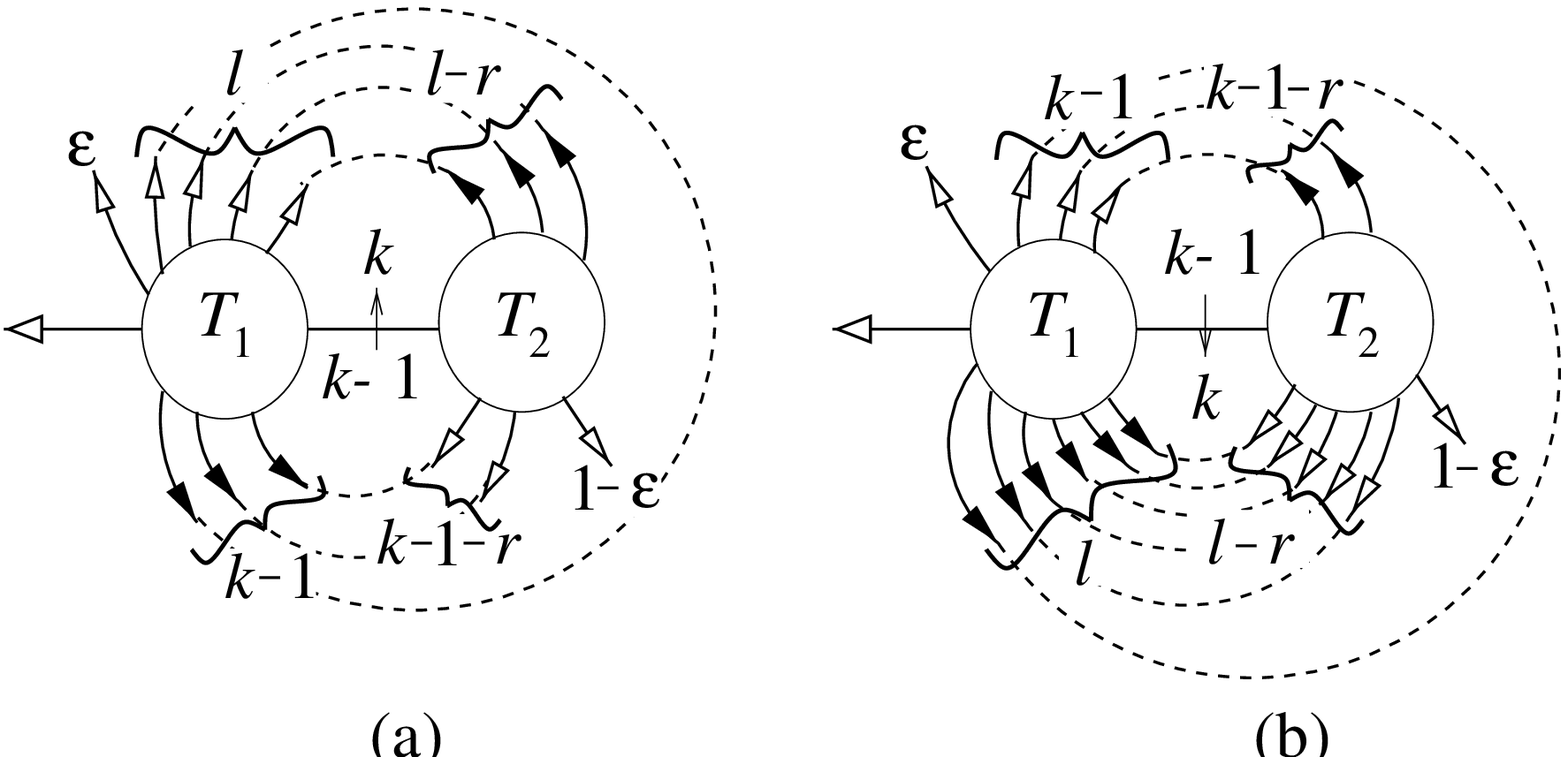}{12.cm}
\figlabel\inv

More precisely, let us start with an unrooted R-tree $T$, and its associated two-leg diagram $\Gamma$
by bud-leaf connection. Let $T'$ further denote the unrooted R-tree obtained by cutting
$\Gamma$: we want to show that $T'=T$. Noting that  
$T$ and $T'$ have the same number of inner edges (=number of inner vertices 
$-$ 1), and therefore the same number of bud-leaf pairs (=number of inner edges 
in $\Gamma$ $-$ number of inner edges in $T$ or $T'$),  
it is sufficient to show that the inner edges of $T$ 
cannot belong to leftmost minimal paths of $\Gamma$ (which would otherwise be cut
into a bud-leaf pair in $T'$).
Assume by contradiction that some leftmost minimal paths of $\Gamma$ contain 
some inner edges of $T$. We pick such a path $\{ e\}$ to the face $f$
with minimal length $k$. 
Obviously $e_1,e_2,...,e_{k-1}$ correspond to bud-leaf pairs of $T$, while
$e_k$ is one of its inner edges. 
Let $T_1$ and $T_2$ 
denote the pieces of $T$ separated upon cutting $e_k$, and such that 
the marked (unmatched) leaf belongs to $T_1$. 
We introduce
as usual the quantity $\epsilon=1$ or $0$
according to whether the other unmatched leaf of $T$ is in $T_1$ or $T_2$.
Let $\ell$ finally denote the ``depth" of $f$ in $T$, namely that of the bud-leaf pair
whose connection closes the face $f$ when gluing $T$ into $\Gamma$. 
The generic situation is depicted in 
Fig. \inv, according to whether $T_1$ lies on the left
(Fig. \inv\ (a)) or right (Fig. \inv\ (b))
when going to $f$ across $e_k$. 
In the case (a), the edges $e_1,e_2,...,e_{k-1}$ originate from buds of $T_1$, 
the first $r$ of which are connected to leaves of $T_1$ (encircling $T_2$), while the remaining 
$k-1-r$ are connected to leaves of $T_2$. Note that $r=0$ necessarily when $\epsilon=0$. 
In the case (b), these edges terminate at leaves of $T_1$, the first $r$ of which
originate from buds of $T_1$ (encircling $T_2$), while the remaining $k-1-r$ originate
from buds of $T_2$ (with again $r=0$ if $\epsilon=0$). 
The total charge of $T_2$ in case (a) reads
\eqn\chartwo{q(T_2)= (1-\epsilon)+(k-r-1)-(\ell-r)= (k-\ell)-\epsilon}
The minimality of $\{ e\}$ implies that $\ell\geq k$, and as $\epsilon \geq 0$,
the charge \chartwo\ is negative or zero, therefore it must be $-1$ (from the condition (UR4')
characterizing unrooted R-trees), but this in turn implies that the vertex on the right
of $e_k$ is occupied, contradicting the orientation constraint of minimal paths.
We  are left with case (b), in which the total charge of $T_1$
reads
\eqn\charone{q(T_1)= 1+\epsilon+(k-1)-\ell=\epsilon+k-\ell }
Again, the minimality condition imposes that $\ell\geq k$, but the leftmost
condition further eliminates the case $\ell=k$, hence $\ell>k$ and as
$\epsilon\leq 1$ the charge 
\charone\ is negative or zero, therefore it must be $-1$ again, and the vertex on the right 
of $e_k$ is occupied, contradicting the orientation constraint of minimal paths. 
In all cases, we have found a contradiction, henceforth  
proving the desired statement that $T=T'$.

\subsec{Enumeration via rooted trees}

By virtue of the bijection established in Sects. 3.1, 3.2 and 3.3 above, the computation of 
$\gnn$ translates into the enumeration of unrooted R-trees. It is a general fact that 
rooted trees are naturally simpler to enumerate, as they can be generated recursively. 
Such trees are naturally oriented from the root to their endpoints, with an
obvious notion of descending subtrees.

\leftline{\sl 3.4.1. Rooted R-trees}
\nobreak
In this section, we first consider rooted R-trees, defined as unrooted R-trees 
with a distinguished leaf replaced by a neutral (charge zero) root. 
A rooted R-trees is entirely characterized by the following properties
\item{(R1)} This tree is a ternary rooted tree with empty and occupied inner vertices, inner edges
obeying the particle exclusion rule and with two types of endpoints: buds and leaves    
\item{(R2)} No leaf is connected to an occupied vertex and the root is connected to an empty
vertex
\item{(R3)} Attaching the charge $+1$ to leaves and $-1$ to buds as well as 
$0$ to the root, the total charge of this tree is $+1$
\item{(R4)} Any descending subtree starting with an occupied vertex
has total charge $+1$ or $-1$, while any descending subtree starting with an 
empty vertex has total charge $+1$ or $+3$, the latter moreover  
descending necessarily from an occupied ancestor vertex 
\par
These properties are clearly implied by the characterization (UR1)-(UR4') of unrooted R-trees.
Conversely, given a tree satisfying (R1)-(R4), we replace the root by a leaf of charge $+1$.
The resulting tree clearly satisfies the properties (UR1)-(UR3). 
Moreover, when cutting an inner
edge, the formerly descending subtree either starts with an occupied vertex
and has charge $+1$ or $-1$ (from (R4)), in which case the complementary part starts
with an empty vertex and has charge $+1$ or $+3$ respectively, or starts
with an empty vertex and has charge $+1$ or $+3$ (from (R4)), in which 
case the complementary part has charge $+1$ or both has charge $-1$ and starts with
an occupied vertex (as ancestor of a descending subtree of charge $+3$).
In all cases, (UR4) is satisfied.

\leftline{\sl 3.4.2. More rooted trees}
\nobreak
\fig{Pictorial representation of the recursive relations (3.3) and (3.4)
obtained by listing all possible environments of the starting vertex in each
type of tree. The introduction of V-trees, namely all rooted trees of 
charge $1$ (i.e single leaves, R- or X-trees), although redundant, is useful
for the compactness of relations (3.4). 
}{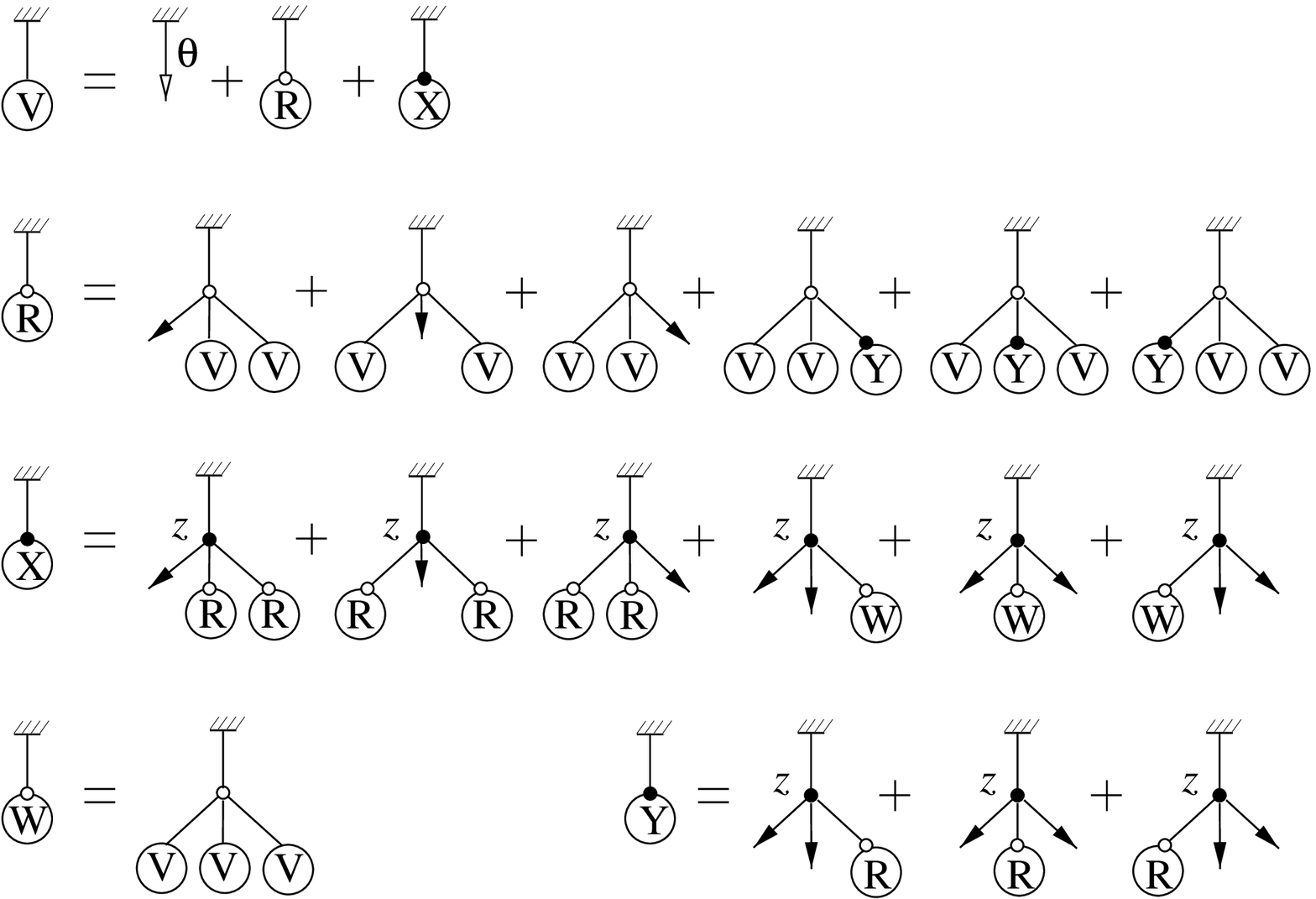}{13.5cm}
\figlabel\rwxytree

As a first consequence of the characterization (R1)-(R4) of rooted R-trees, 
we note that their descending subtrees of charge $+1$ starting with an empty vertex
are themselves rooted R-trees. The three other types of subtrees are
rooted trees of a different nature, which will be called W-trees (charge $+3$, starting from an
empty vertex), X-trees (charge $+1$, starting from an
occupied vertex) and Y-trees (charge $-1$, starting from an
occupied vertex). As for rooted R-trees, these other trees are all characterized by similar
properties, respectively (W1-W4), (X1-X4), (Y1-Y4), where (W1)=(X1)=(Y1)=(R1), 
(W4)=(X4)=(Y4)=(R4),  with the obvious modifications of (R2)
for the nature of the vertex attached to the root namely (W2)=(R2) and
\item{(X2)}=(Y2)  No leaf is connected to an occupied vertex and the root is connected to an 
occupied vertex
\par
\noindent and with the obvious modifications of
(R3) for the total charge, namely (X3)=(R3) and
\item{(W3)}  Attaching the charge $+1$ to leaves and $-1$ to buds as well as
$0$ to the root, the total charge of this tree is $+3$
\item{(Y3)} Attaching the charge $+1$ to leaves and $-1$ to buds as well as
$0$ to the root, the total charge of this tree is $-1$   
\par

\leftline{\sl 3.4.3. Generating functions}
\nobreak
All descending subtrees of R,W,X,Y-trees not reduced to single buds or leaves
are themselves R,W,X,Y-trees. We introduce the generating functions 
$R,W,X,Y$ for the corresponding objects, with now a weight $\theta$ 
per {\it leaf} and $z$ per occupied vertex. 
Moreover, it proves convenient to introduce the function
\eqn\Vdef{ V=\theta+R+X}
which generates V-trees, namely {\it all possible subtrees of charge $+1$} 
(including single leaves).

The generating functions for our trees satisfy the following recursive relations
\eqn\rwxy{\eqalign{
R&=3 V^2 + 3 V^2 Y \cr 
W&=V^3\cr
X&= 3 z R^2+3 z W \cr
Y&= 3 z R\cr} }
obtained by listing all possible environments of the starting vertex of each type of tree,
as depicted in Fig. \rwxytree.
Eliminating $W,X,Y$, we end up with the equation \master\ relating $R$, $V$, $\theta$ and $z$.
This allows therefore to interpret the function $R$ of eqn. \master\ (as obtained from 
Ref. \HARD) as the generating function for the rooted R-trees 
and $V$ as that of the V-trees defined above. 

\leftline{\sl 3.4.4. Local environments within unrooted R-trees}
\nobreak
\fig{A listing of all possible local environments in an unrooted R-tree 
for (a) an occupied vertex, (b) an empty vertex and (c) an inner edge.
In (a), the first three cases differ by the cyclic order of the subtrees
around the occupied vertex at hand.}
{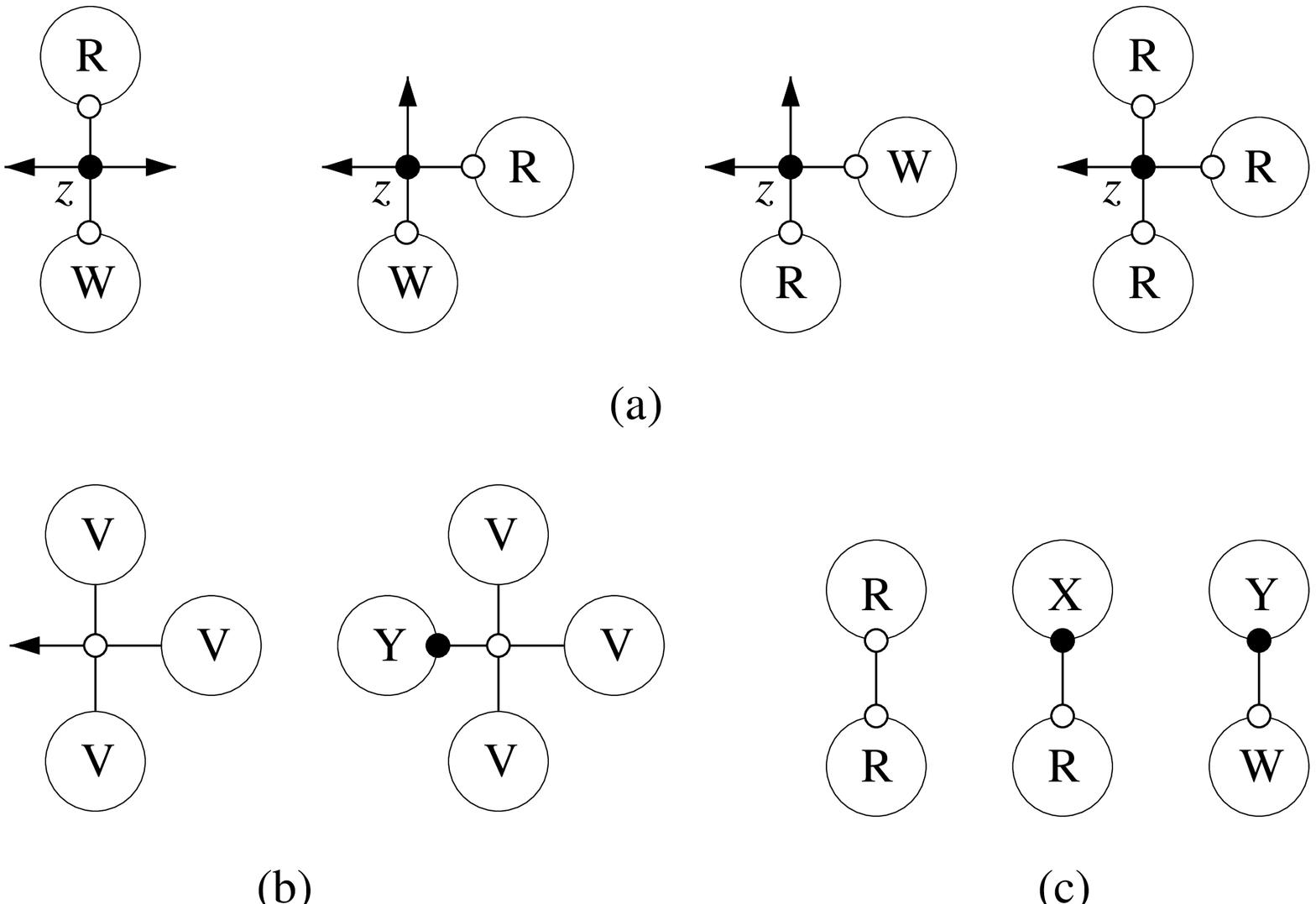}{10.cm}
\figlabel\environ 

To make an even tighter connection between on one hand unrooted R-trees
and on the other hand all rooted trees defined above, we note that 
these rooted trees are precisely the pieces obtained when reading an
unrooted R-tree in any direction from one of its inner vertices 
or inner edges. This property allows to characterize the local 
environment of any inner edge or vertex within an unrooted R-tree,
as depicted in Fig. \environ. It is easily seen that properties (UR1)-(UR4')
imply that only the depicted cases are obtained. Conversely, given one
of these local environments, the resulting tree clearly obeys (UR1)-(UR3),
and (UR4') may be checked by considering one of its inner edges, say $e$.
This edge may either belong to the local environment itself (edges
depicted in the figure) or be within one of the rooted subtrees.
In the first case, (UR4') is directly apparent from the figure,
while in the second case, one of the pieces originating from $e$ is 
a subtree of the rooted tree at hand, and we use the characterization 
(R1)-(R4) (and their modifications) for all rooted trees to check
that (UR4') is satisfied. 

\leftline{\sl 3.4.5. Computation of $\gnn$}
\nobreak
So far we have obtained the generating functions for rooted trees. 
Let us now make the connection with the generating function 
of unrooted R-trees, and eventually with $\gnn$.
We note first that the weight $\theta$ per leaf in
a rooted tree of charge $q$ is equivalent to a weight $\theta$ per 
inner vertex times an overall weight $\theta^{(q+1)/2}$ 
by virtue of Euler's relation for tetravalent graphs. 
In the case of rooted R-trees ($q=1$), this coincides precisely with our previous 
convention for $\gnn$ with a weight $\theta$
per inner vertex, times a weight $\sqrt{\theta}$ for each of the two external legs.

To identify the generating function $\gnn$, recall that the two leg diagrams with both legs
occupied are in one-to-one correspondence 
with unrooted R-trees with a marked leaf at depth $0$.
The depth 0 leaves are precisely those not matched with a bud, hence
$\gnn$ is equal to the difference between the generating function of
unrooted R-trees with a marked leaf (i.e. rooted R-trees) and that of 
unrooted R-trees with a marked bud. The former generating function
is nothing but $R$. Examining
all possible local environments of buds in Fig. \environ, the latter
generating function reads $(V^3+zR^3+6zRW)/\theta$, where the overall factor
$1/\theta$ has been tuned to match the weight prescription for $\gnn$. 
Eqn. \gamtwo\ follows.

As a side remark, we note that the relation between unrooted R-trees with a marked leaf at depth 
$0$ and rooted R-trees is $2$-to-$(n+1)$,
where $n$ denotes the number of leaves in the rooted R-tree. Indeed, 
in the unrooted R-tree, each of the $(n+1)$ leaves may serve as roots for a rooted R-tree,
but only two of them have $0$ depth\foot{
In the case of an accidental half-turn symmetry of the unrooted R-tree (which is the only
possible symmetry), the correspondence is rather one-to-$(n+1)/2$.}.  
This is a manifestation of the notion of {\it conjugacy} of trees introduced in Ref. \SCH,
which translates into the relation
\eqn\gamtoR{ \gnn\vert_{\theta^n} = {2\over n+1} \ R\vert_{\theta^n} \quad n\geq 2 }
where the subscript $\theta^n$ refers to the coefficient of $\theta^n$ in
the corresponding functions. This relation can be also verified directly
by checking that $d(\theta\gnn)/d\theta=2R$, using eqns. \gamtwo\ and \master.

\newsec{Four-leg diagrams}

We now turn to the enumeration of four-leg diagrams contributing to 
$\gnnnn$ by adapting the cutting procedure defined above.

We start with a four-leg diagrams with occupied external legs labeled
by, say, 1, 2, 3 and 4 in counterclockwise cyclic order. 
Note that such a four-leg diagram may be either connected, or made
of two (connected) two-leg diagrams. In the latter case, the connected
end-points are either (1-2) and (3-4) or (1-4) and (2-3).
This translates into the relation
\eqn\connec{\gnnnn=\gnnnnc+2\left(\gnn\right)^2}
where $\gnnnnc$ denotes the generating function for {\it connected}
four-leg diagrams with occupied external legs. From now on, we
will consider the case of connected diagrams only.

\subsec{Cutting procedure}

\fig{The cutting of the two-leg diagram (with external legs 1 and 4)
constructed by gluing the legs 2 and 3 of a connected four-leg 
diagram into an edge $e$. This edge is either cut and replaced by a bud-leaf pair (a)
or remains uncut (b). In the latter case, it separates the tree resulting from the cutting
into two pieces $T_1$ and $T_2$ where: $T_1$ contains both legs 1 and 4, exactly one
bud of $T_1$ is connected to a leaf of $T_2$ as shown, and no bud of $T_2$ is
connected to a leaf of $T_1$.}{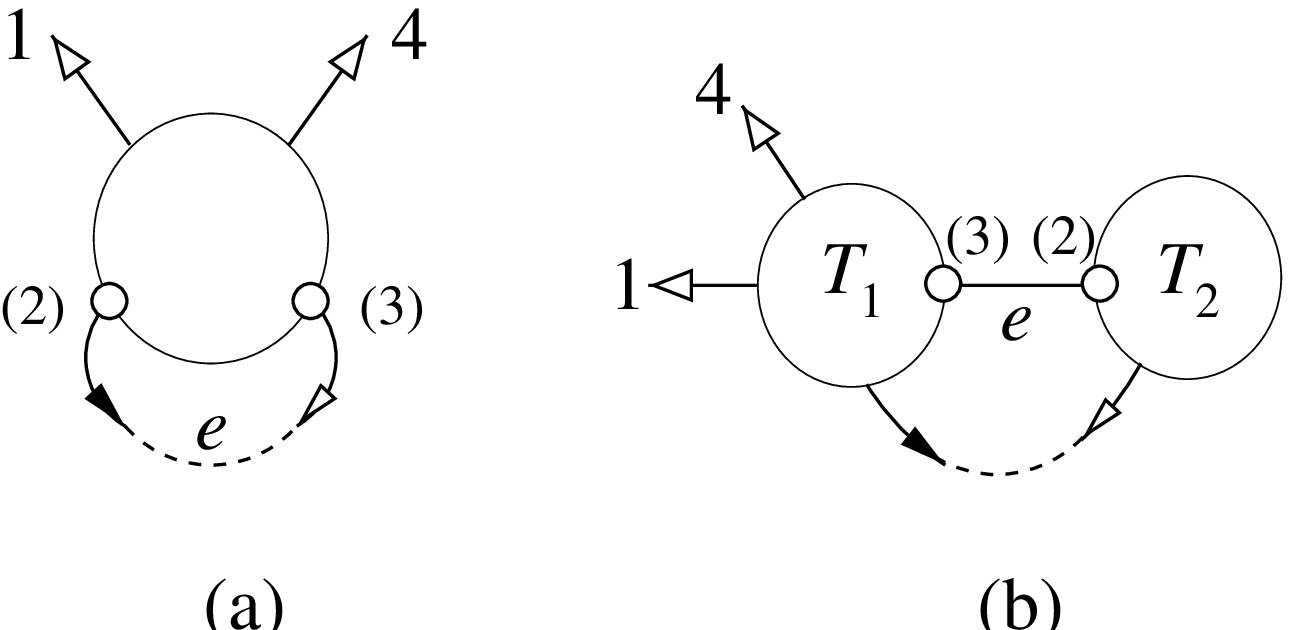}{9.cm}
\figlabel\gamfourcut

To use the cutting procedure of the previous section, we first
connect the legs 2 and 3, erasing the associated occupied univalent 
vertices. The result is a two-leg diagram with external legs 1 and 4
and with a distinguished edge such that 
\item{(i)} it connects two inner white vertices
\item{(ii)} it is adjacent {\it on one side only} to the external face 
\item{(iii)} it lies between 1 and 4 in counterclockwise order around 
the external face.
\par
The correspondence is clearly one-to-one. Now we proceed by applying
the cutting procedure to the two-leg diagram, choosing 1 for the incoming
leg. Two situations may occur: (a) the distinguished edge is cut and replaced
by a bud-leaf pair, or (b) it remains uncut. These situations are
depicted in Fig. \gamfourcut. 

In case (a), we can transfer the marking of the distinguished edge 
into the marking of the associated bud. We end up with an unrooted
R-tree with a distinguished bud connected to a white vertex and 
of depth 1. We denote by $\Rbudb{1}$ the generating function for such objects. 
Conversely, given an unrooted R-tree with such marking,
the gluing procedure clearly gives rise to a two-leg diagram with
a marked edge with properties (i) as the leaf connected to
the marked bud necessarily originates from a white vertex from (UR2),
(ii) from the depth 1 criterion, and (iii) by choosing the incoming
and outcoming legs 1 and 4 appropriately among the two leaves of depth 0.

In the case (b), the distinguished edge separates the resulting unrooted
R-tree into two pieces $T_1$ and $T_2$, with $T_1$ containing the former
external leg $1$. From condition (i) for the distinguished edge, 
those pieces are rooted R-trees as shown in Fig. \environ (c). Moreover, 
the condition (ii) ensures that there is exactly one bud in
$T_1$ matched to a leaf in $T_2$, and no bud in $T_2$ is
matched to a leaf in $T_1$. The charge constraint for rooted R-trees shows
that the two unmatched leaves both lie in $T_1$, in the position
depicted in Fig. \gamfourcut\ (b). Cutting finally the distinguished edge 
and replacing it by a leaf on each side, we end up we two unrooted R-trees, one
with a marked leaf at depth 1 (corresponding to $T_1$) and one
with a marked leaf at depth 0 (corresponding to $T_2$). 
The generating function of these objects is $\Rleaf{0}\times\Rleaf{1}$,
where $\Rleaf{k}$ denotes the generating function of unrooted R-trees
with a marked leaf of depth $k$.  
Conversely,
given two such marked trees and connecting the marked leaves into an edge, 
we get an unrooted R-tree with a marked edge which upon the gluing 
procedure produces a two-leg diagram with a distinguished edge satisfying
(i),(ii) and (iii) by choosing the incoming and outcoming legs 1 and 4 
appropriately.

We end up with the following relation:
\eqn\gamfourrel{\gnnnnc= \Rbudb{1}+ \Rleaf{0}\times\Rleaf{1}}
The computation of $\gnnnnc$ therefore reduces to the enumeration of
unrooted R-trees with suitable markings.

\subsec{Computation of $\Rbudb{1}$ via rooted trees} 

\leftline{\sl 4.2.1. Unrooted W-trees} 
\nobreak
We now turn to the precise calculation of $\Rbudb{1}$. As a preliminary
remark, we note that replacing the marked bud by a root in any tree
contributing to $\Rbudb{1}$ yields a rooted W-tree as apparent from  
Figs. \rwxytree\ and \environ\ (b), since the marked bud is by definition
connected to an empty vertex. 
Replacing the root by a leaf produces an {\it unrooted} W-tree of total charge 
$+4$ with a marked leaf.  
The characterization of unrooted W-trees is somewhat more subtle than that
(UR1-UR4) of unrooted R-trees. In particular, not all the leaves of the
unrooted tree may lead, when replaced by roots, to a rooted W-tree. 
We call {\it admissible} the leaves actually leading to rooted W-trees.
The set of admissible leaves will be characterized precisely in the next section.

The depth $1$ constraint on the formerly
marked bud now translates into the fact that the (admissible) marked leaf in the unrooted
W-tree has depth $0$ when performing the bud-leaf matching directly in the unrooted W-tree. 
Note that in an unrooted W-tree there are exactly four depth $0$ leaves.
Unfortunately, in general, not all of them are admissible.   
We may still write 
\eqn\rbubble{ \Rbudb{1}=0\times {\cal W}^{(0)} +1\times  {\cal W}^{(1)}+2 \times {\cal W}^{(2)}+
3 \times {\cal W}^{(3)}+4 \times {\cal W}^{(4)}}
where ${\cal W}^{(i)}$, $i=0,1,2,3,4$, is the generating function for unrooted W-trees with
exactly $i$ admissible leaves of depth $0$ (necessarily connected to an empty vertex
as part of the definition of W-trees).  

\leftline{\sl 4.2.2. Admissible leaves and the core of unrooted W-trees}
\nobreak
As a tool to characterize admissible leaves,
we now come to the definition of the {\it core} of an unrooted W-tree.  
Starting from an unrooted W-tree, let us  
\item{(i)} first mark all edges separating
the tree into pieces of charge $+1$ and $+3$, in which the piece of charge $+1$
starts with an empty vertex
\item{(ii)} then cut all these marked edges and replace them
by a leaf (on the charge $+3$ side) and a bud (on the other side)
\par

\fig{The core of an unrooted W-tree is obtained by cutting all edges adjacent
to a piece of charge $+1$ starting with an empty vertex (indicated by parallel double-lines).
Replacing these cut edges by bud-leaf pairs with a leaf on the charge $+3$
complementary side,
the core is the only connected component with total charge $+4$. It is indicated here in thick
lines. The pieces connected to the core are either leaves, buds or
maximal R-subtrees (circled in the figure).}{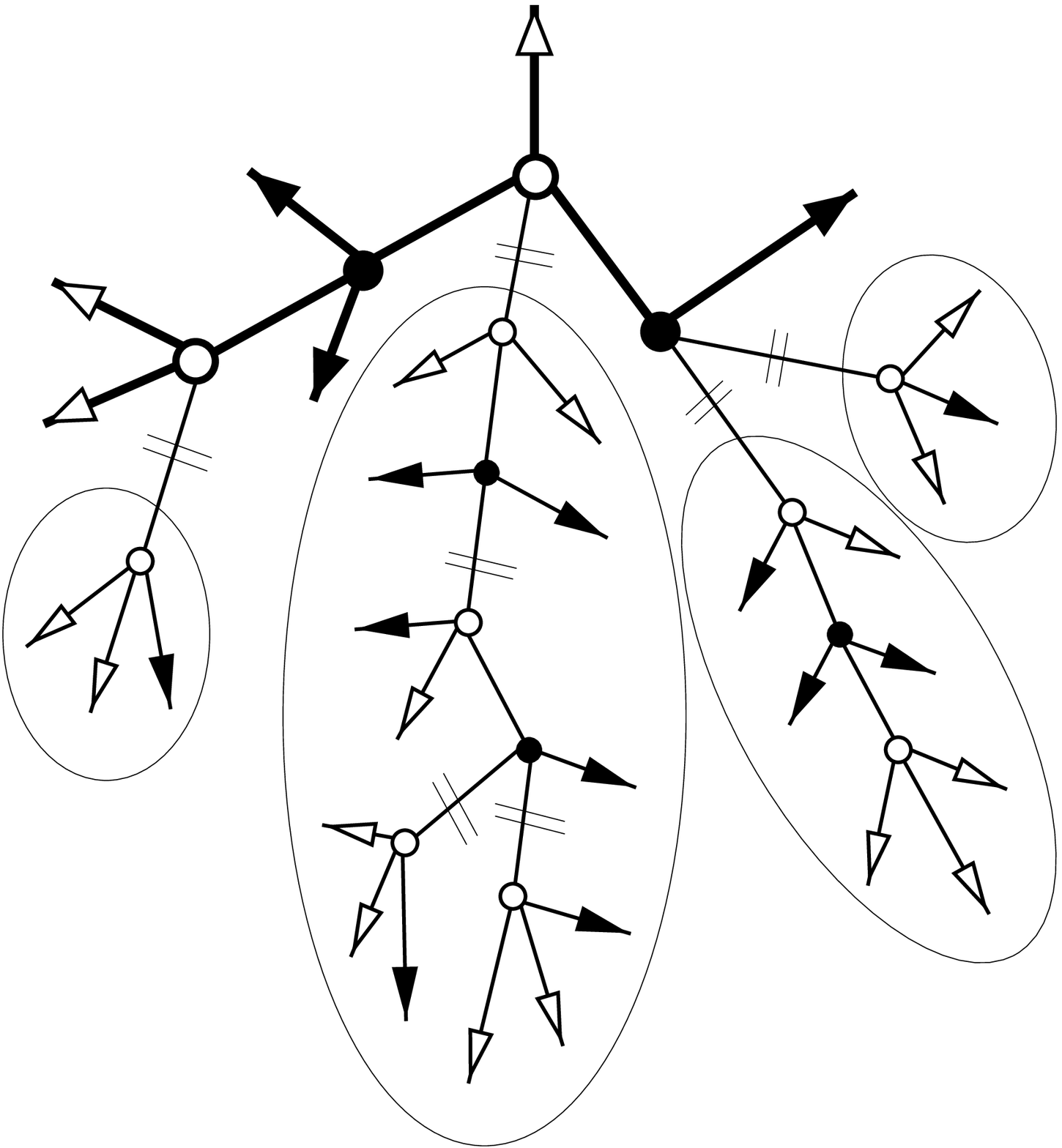}{8.cm}
\figlabel\wcore

The resulting object is a set of disconnected pieces, all of which have charge $0$
except one, which has charge $+4$, and which we will call the core of the unrooted W-tree.
Indeed, picking an admissible leaf and replacing it by a root, the edges marked at
step (i) above are the root edges of descendent R-subtrees, as there are no descendent
subtrees of charge $+3$ with an empty ancestor vertex in a rooted W-tree.
At step (ii) above, all these descendent R-subtrees acquire a charge $0$ and are
amputated from their own R-subtrees, acquiring a charge $0$ as well. The only
remaining connected component is that
containing the selected admissible leaf of the unrooted W-tree and has total charge $+4$.
As the procedure (i)-(ii) was defined independently of a choice of admissible leaf at hand,
this shows moreover that all admissible leaves lie in the core. 

It remains to show the converse, namely that all the leaves of the core are admissible.
Note first that all trees not reduced to buds or leaves attached to the core are rooted R-trees,
which we call {\it maximal} R-subtrees. 
Let us show that all inner edges of the core separate the tree into a piece of charge $+3$
starting with an empty vertex, and one of charge $+1$ 
starting with an occupied one. 
One of these two pieces is a descendent subtree of the rooted W-tree hence as such
is either an X-subtree or a W-subtree as R-trees have been cut out from the core
and Y-trees only occur as subtrees of maximal R-subtrees as
readily seen in Fig. \rwxytree.
The other piece has the complementary charge ($+3$ and $+1$ respectively) 
and starts with an empty vertex in the first case according to the 
particle exclusion rule, or with an occupied vertex in the second case as
the ancestor of a W-subtree. 
We are now ready to show that replacing a leaf in the core by a root
produces a rooted tree obeying (W1-W4).
Properties (W1-W3) are obviously satisfied. To show (W4), consider any
descendent subtree not made of a single bud
or leaf. Two situation may occur.
Either it is contained in an R-subtree, 
in which case we use the property (R4) of the R-subtree to ensure that
the subtree at hand meets the criteria of (W4). Or it 
originates from an edge in the core, 
in which case we use the above property of the inner edges of the core 
to again meet the criteria of (W4).

To conclude this section,  
the unrooted W-trees may be viewed as cores
to which are attached leaves, buds or maximal R-subtrees (see
Fig. \wcore\ for a typical example). When replacing the maximal
R-subtrees by a new type of endpoint which we call R-{\it leaves}, the cores of unrooted
W-trees satisfy the following properties
\item{(C1)} These are trees made of tetravalent empty or occupied inner vertices,
and three kinds of endpoints: leaves, buds and R-leaves
\item{(C2)} No leaf is connected to an occupied vertex
\item{(C3)} Attaching a charge $+1$ to leaves and R-leaves and $-1$ to buds, the total
charge is $+4$
\item{(C4)} Cutting any inner edge separates the cores into two pieces: one of charge $+3$ 
starting with an empty vertex, and one of charge $+1$ starting with an occupied vertex
\par
In particular, (C4) is stronger than the hard-particle exclusion as it
ensures that empty and occupied vertices alternate (bipartite tree).

\leftline{\sl 4.2.3. Unrooted C-trees}
\nobreak
The properties (C1-C4) of previous section define a new type
of unrooted trees obtained by replacing R-leaves by arbitrary rooted R-trees.
These objects are called unrooted C-trees.
These are more general than unrooted W-trees.
Indeed, 
our construction of unrooted W-trees from rooted ones implicitly assumes
the existence of admissible leaves, which in turn implies that the core 
contains at least one leaf.  
This constraint is not implied by (C1-C4), and an unrooted C-tree may have
no leaf in its core. The absence of such a constraint in unrooted C-trees 
makes them easier to deal with.
We may rewrite eqn. \rbubble\ in terms of C-trees as well upon introducing
the generating functions ${\cal C}^{(i)}$, $i=0,1,2,3,4$ for unrooted 
C-trees with exactly $i$ depth $0$ leaves in the core, namely
\eqn\cbubble{  \Rbudb{1}=0\times {\cal C}^{(0)} +1\times  {\cal C}^{(1)}+2 \times {\cal C}^{(2)}+
3 \times {\cal C}^{(3)}+4 \times {\cal C}^{(4)}}
as we clearly have ${\cal C}^{(i)}={\cal W}^{(i)}$ for $i\neq 0$.

As in the case of unrooted W-trees, the notion of core of an unrooted C-tree is unambiguous
as it can be recovered from the whole tree (completed with maximal R-subtrees) by
the procedure (i)-(ii) of Subsect. 4.2.2. In particular, the enumeration of 
unrooted C-trees will boil down to that of their cores.   

\leftline{\sl 4.2.4. Enumeration of unrooted C-trees}
\nobreak
\fig{Local environment {\it in the core} of an unrooted C-tree for (a) an empty vertex 
(b) an occupied vertex (c) an inner edge. In case (a) we have decomposed one of the
attached V-trees to display a leaf and a maximal R-subtree.}{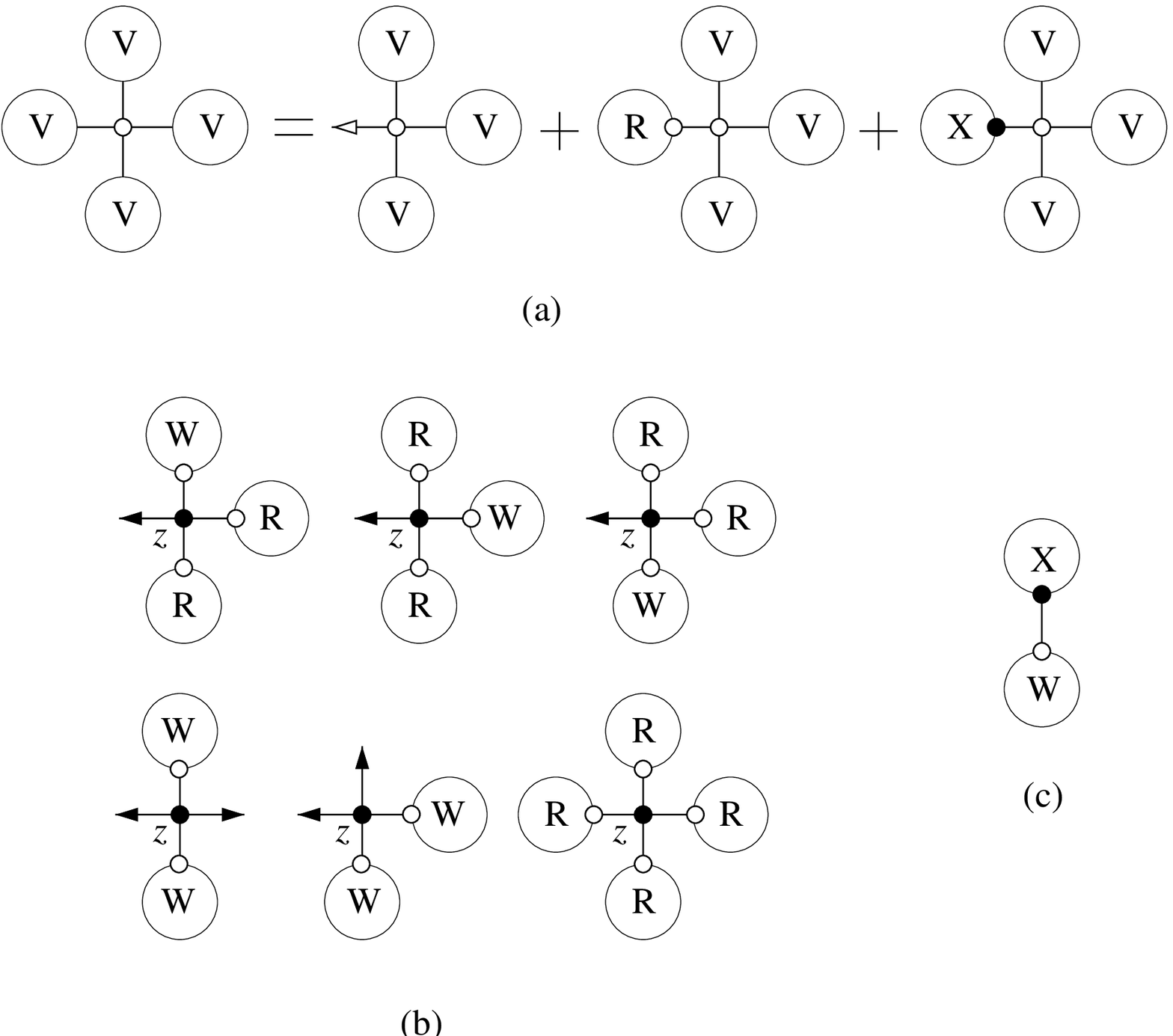}{13.cm}
\figlabel\cenviron

The enumeration  of unrooted C-trees is performed by noting that in the core there are
four more leaves or R-leaves than buds. The generating function ${\cal C}$ of unrooted C-trees 
reads
\eqn\cbub{ {\cal C}={1\over 4}\left( \Cleaf + \Crleaf - \Cbud\right)}
where $\Cleaf,\Crleaf,\Cbud$ denote the generating function
for unrooted C-trees with respectively a marked leaf in the core, a distinguished
maximal R-subtree and a marked bud in the core.

These are computed by examining the local environment of a vertex 
in the core of an unrooted C-tree, depicted in Fig. \cenviron\ (a-b).
One may easily check that properties (C1-C4) imply that only the depicted cases 
are possible. Conversely, given one of these local vertex environments, we construct the
core by applying the procedure (i)-(ii) of Subsect. 4.2.2: let us consider an edge $e$ cut
in the procedure and show that the piece not containing the vertex at hand is
an R-subtree. 
The edge $e$ may either be connected to the  vertex itself
(edge depicted in Fig. \cenviron) or be within one of the rooted subtrees depicted.
The first case is easily worked out by inspecting Fig. \cenviron\ (a-b), while the second
case requires the use of the characterizations (R1-R4) for R-trees and their modified versions
for W,X,Y-trees to show that an R-subtree not containing the vertex at hand
is cut out. 
As in Subsect. 4.2.2, we end up with cut out pieces of charge $0$, and a unique
piece of charge $+4$, containing the vertex at hand, which is the core. 
The properties (C1-C4) follow immediately. 

By inspection in Fig. \cenviron\ of all possible environments of leaves, maximal R-subtrees and
buds in the core of an unrooted C-tree, we find
\eqn\variousc{\eqalign{
\Cleaf &= V^3 = W \cr
\Crleaf &= {R (V^3+6z RW +z R^3) \over \theta} \cr
\Cbud &= {3 z W^2+ 3 z R^2 W\over \theta} \cr}}
leading to
\eqn\ccalc{ {\cal C}={1\over 4}\left( V^3 + {R\over \theta}
(V^3+6zRV^3+zR^3) -{3z V^6+3z R^2V^3\over \theta} \right) }

\leftline{\sl 4.2.4. Calculation of $\gnnnnc$ }
\nobreak
In terms of the ${\cal C}^{(i)}$, the generating function ${\cal C}$ is expressed
as ${\cal C}=\sum_{i=0}^4 {\cal C}^{(i)}$, hence using 
eqns \gamfourrel\ and \cbubble, we have
\eqn\gafin{ \gnnnnc= 4 {\cal C} -\left( \sum_{i=0}^4 (4-i) {\cal C}^{(i)} - \Rleaf{0}
\Rleaf{1} \right) }
We will now show that 
\eqn\resapp{  \sum_{i=0}^4 (4-i) {\cal C}^{(i)} - \Rleaf{0}
\Rleaf{1} = {2 \over \theta^2} (V^3+6zRV^3+zR^3)^2 }
by identifying $\sum_{i=0}^4 (4-i) {\cal C}^{(i)}$ with the generating function
of unrooted C-trees with a marked leaf of depth $0$ not in the core i.e.
in a maximal R-subtree, and by examining
the resulting constraints on this R-subtree. 
Together with eqns. \ccalc\ and \gafin, this finally yields
\eqn\gfourconnex{ \gnnnnc= V^3 +
{RV^3+3zR^2V^3+zR^4 -3z V^6\over \theta} -2 {(V^3+6zRV^3+zR^3)^2 \over \theta^2} }  
and eqn. \gamfour\ follows from eqn. \connec. 

To prove eqn. \resapp, we first note that the quantity
$\sum_{i=0}^4 (4-i) {\cal C}^{(i)}$ enumerates unrooted C-trees with (i) a distinguished
maximal R-subtree and (ii) a marked depth $0$ leaf within this subtree.
Introducing the generating
functions ${\Crleaf}_i$, $i=0,1,2,3,4$ for unrooted C-trees with a distinguished
maximal R-subtree containing exactly $i$ depth $0$ leaves,
we have
\eqn\clearly{ \sum_{i=0}^4 (4-i) {\cal C}^{(i)} =\sum_{i=0}^4 i\, {\Crleaf}_i }

We will now proceed in two steps. In a first step we characterize unrooted C-trees with a
distinguished maximal R-subtree, and in a second step we exhaust all possible cases according
to the number of depth $0$ leaves in the maximal R-subtree.  

Starting from an unrooted C-tree with a distinguished maximal R-subtree,
we cut the edge connecting it to the core, and replace it by a leaf on the side of
the R-subtree and a bud on the side of the core. We have clearly on the first side an unrooted
R-tree with a marked leaf, while on the other side we get an unrooted R-tree with a marked bud.
This last point is readily seen by replacing in Fig. \cenviron\ any R-subtree by a bud
and checking that the resulting environment is one of Fig. \environ, characterizing
unrooted R-trees.  Conversely, given any two unrooted R-trees with respectively
a marked leaf and a marked bud, connecting them by their marked ends yields
an unrooted C-tree whose core contains the vertex attached to the former
marked bud: this is again checked by inspecting the possible local environments 
of this vertex in Figs. \environ\ (a-b) and \cenviron\ (a-b).

\fig{Schematic representation of unrooted R-trees with respectively 
a marked leaf of depth $k$ (contributing to $\Rleaf{k}$)
and a marked bud of depth $l$ (contributing to $\Rbud{l}$).
Gluing the marked leaf and bud into an edge
leads to an unrooted C-tree with a distinguished maximal R-subtree.
The number of depth $0$ leaves in this R-subtree (left piece) 
is (a) $0$ if $k\leq l-2$,
(b) $1$ if $k=l-1$ (c) $2$ if $k=l$ (d) $3$ if $k=l+1$ and (e) $4$
if $k \geq l+2$.}{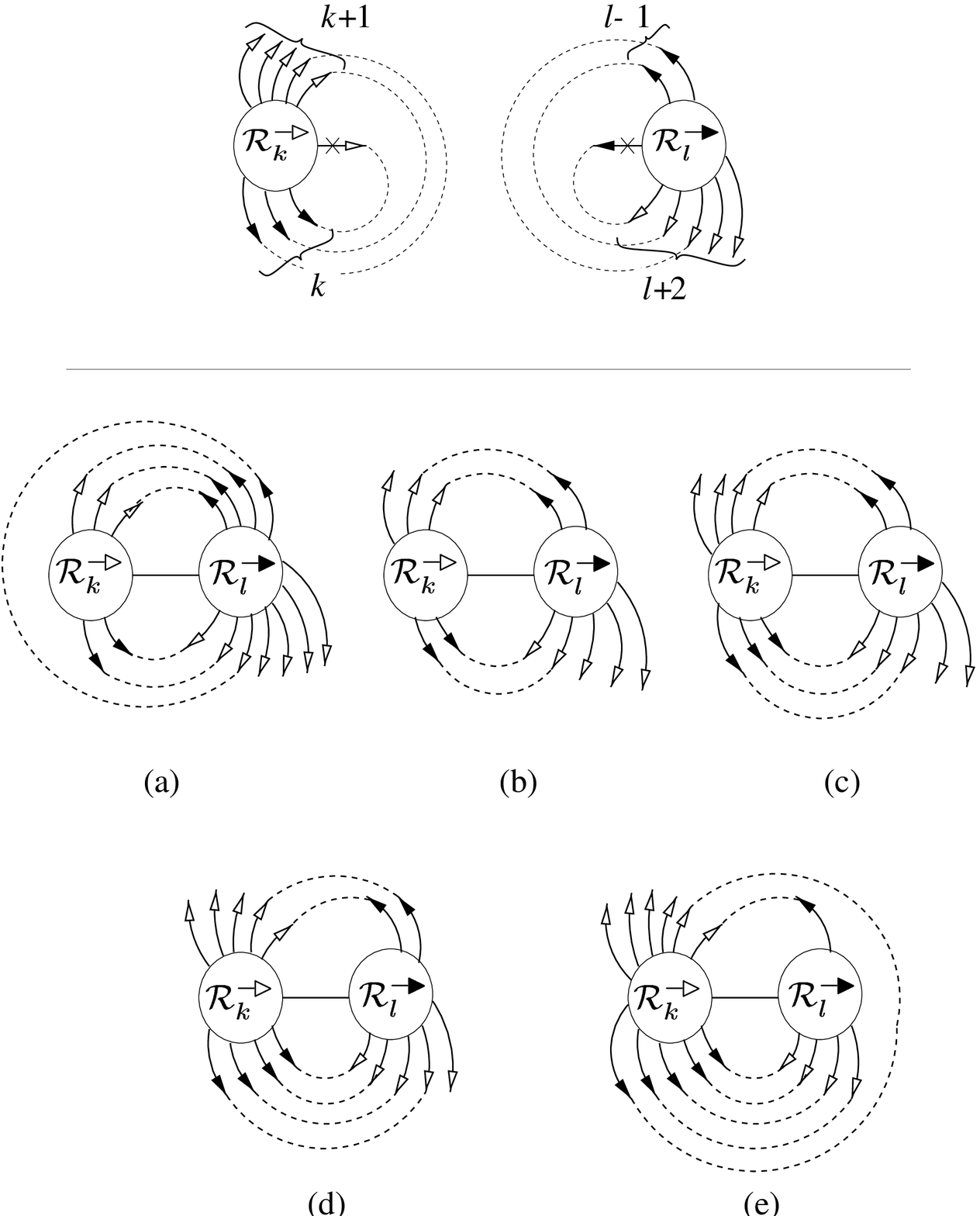}{11.cm}
\figlabel\split

We are now ready to compute the number of depth $0$ leaves in the distinguished 
maximal R-subtree. 
Given an unrooted C-tree with a distinguished maximal R-subtree,
that we split as above
into two marked unrooted R-trees, the number $i$ of depth $0$ leaves in the 
distinguished maximal R-subtree is determined by 
the respective depths of the marked leaf and bud in the two unrooted R-trees.  
The list of all possible cases is depicted in Fig. \split\  and translates
into the following equations
\eqn\rrbar{\eqalign{
{\Crleaf}_0 &= \sum_{k\leq l-2} \Rleaf{k} \Rbud{l} \cr 
{\Crleaf}_1 &= \sum_{k =l-1} \Rleaf{k} \Rbud{l}  \cr 
{\Crleaf}_2 &= \sum_{k =l} \Rleaf{k} \Rbud{l}  \cr   
{\Crleaf}_3 &= \sum_{k =l+1} \Rleaf{k} \Rbud{l}  \cr   
{\Crleaf}_4 &= \sum_{k\geq l+2} \Rleaf{k} \Rbud{l}  \cr }}
where $\Rleaf{k}$ and $\Rbud{l}$ denote the generating functions
for unrooted R-trees with respectively a marked leaf of depth $k$ and a marked bud
of depth $l$. These two functions are related, as leaves of depth $k\geq 1$
are matched bijectively with buds of the same depth, hence $\Rleaf{k}=\Rbud{k}$
for $k\geq 1$. When $k=0$, $\Rbud{0}=0$, while $\Rleaf{0}=\gnn$ as a consequence
of the bijection established in Sect. 3.
This allows for a drastic simplification of eqn. \clearly\
\eqn\sumover{\eqalign{
\sum_{i=0}^4 i\, {\Crleaf}_i &= \sum_{l\geq 1} \Rleaf{l-1} \Rleaf{l} +
2  \sum_{l\geq 1}  \big(\Rleaf{l}\big)^2 \cr
&+ 3  \sum_{l\geq 1}  \Rleaf{l+1} \Rleaf{l}   
+4 \sum_{l\geq 1}\sum_{m\geq 2} \Rleaf{l+m}  \Rleaf{l} \cr
&=2 \big(\sum_{l\geq 1} \Rleaf{l}\big)^2 +\Rleaf{0}\Rleaf{1} \cr}}
where the extra term $\Rleaf{0}\Rleaf{1}$ comes from the $l=1$
term in the first sum.
Noting finally that $\sum_{k\geq 0} \Rleaf{k} =R$, we have
$\sum_{l\geq 1} \Rleaf{l}= R -\gnn=(V^3+6zRV^3+zR^3)/\theta$ from eqn. \gamtwo, and
eqn. \resapp\ follows.

\newsec{Empty-occupied duality} 

\fig{Schematic representation of diagrams contributing to
$\gbb$, $\gbbbb$ and $\gnb$. Upon erasing the empty vertex in the second term
contributing to $\gnb$, eqn. (5.1) follows.}{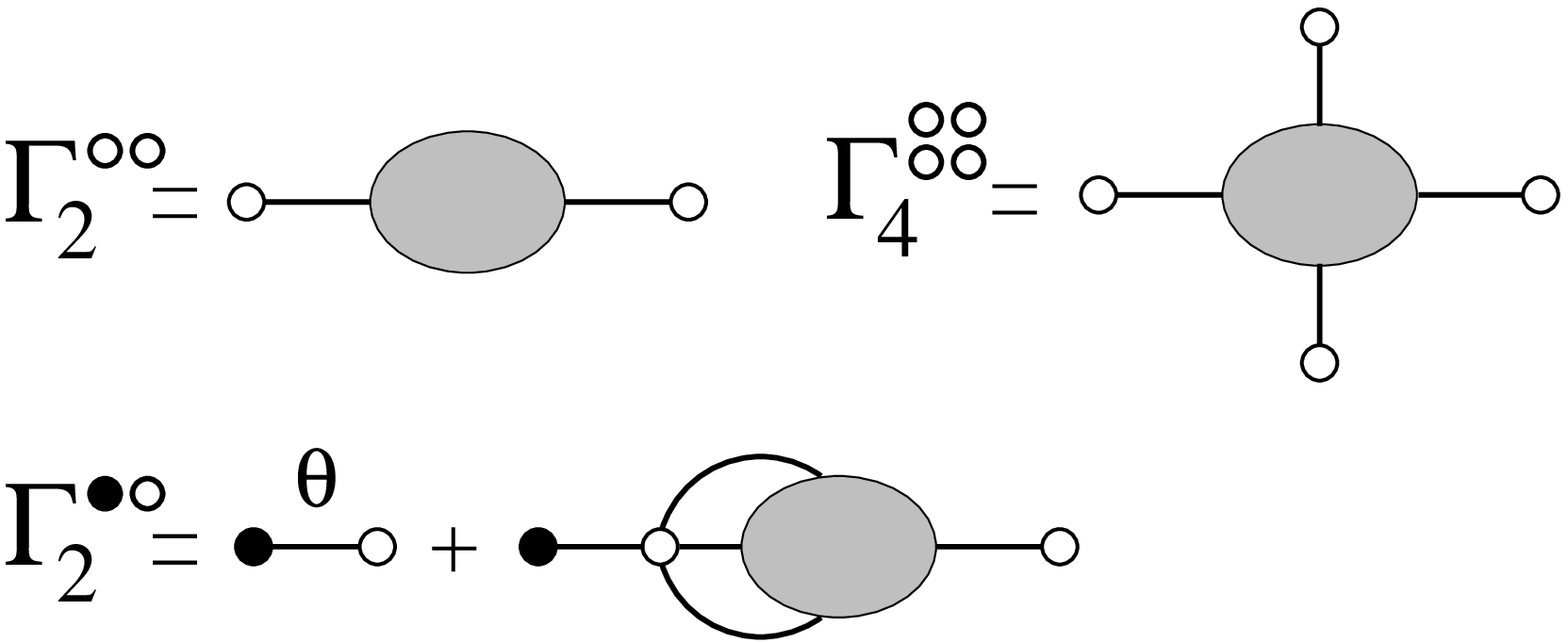}{7.cm}
\figlabel\blacknwhite

\fig{Pictorial representation of the recursive relations (5.2).}{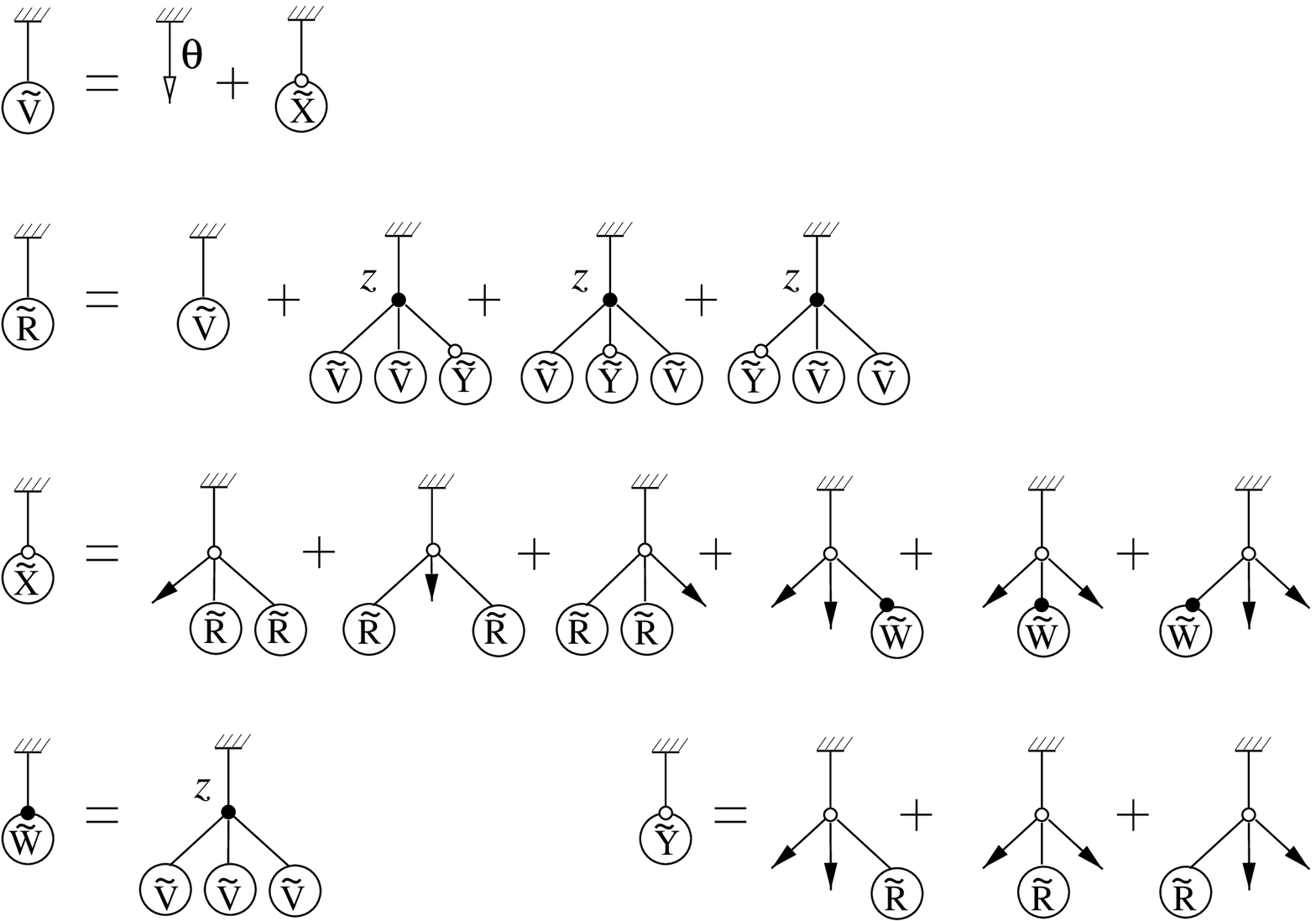}{13.5cm}
\figlabel\white

So far we have calculated $\gnn$, $\gnnnn$, hence $\gnb$ via eqn. \twofour,
which is enough to calculate $E$ via eqn. \energ.
Other quantities of interest are $\gbb$ and $\gbbbb$
counting respectively the two- and four-leg diagrams
with now {\it empty} external legs.  
The latter is easily identified with 
\eqn\relatgb{ \gbbbb = \gnb - \theta}
as depicted in Fig. \blacknwhite.
There is however no such simple expression for $\gbb$. 
To compute it, we note that we can implement a {\it different}
(dual) iterative cutting   
algorithm to produce a tree by changing the rule (ii) of Subsect. 3.1.1
into 
\item{(ii')} the {\it previously visited} vertex is empty 
\par
\noindent This creates a new type of trees where no bud is connected to an
occupied vertex called unrooted ${\tilde {\rm R}}$-trees.
Repeating the analysis of Sect. 3, we are also  
naturally led to consider new types of rooted trees 
called ${\tilde {\rm R}},{\tilde {\rm V}},{\tilde {\rm W}},{\tilde {\rm X}},{\tilde {\rm
Y}}$-trees with the same charges as their untilded counterparts, but with the opposite 
empty/occupied nature of the starting vertices for ${\tilde {\rm W}},
{\tilde {\rm X}},{\tilde {\rm Y}}$-trees as opposed to W,X,Y-trees. 
The generating functions for these new trees satisfy the relations
\eqn\whiterel{\eqalign{
{\tilde V}&= \theta +{\tilde X} \cr
{\tilde R}&= {\tilde V} + 3 z {\tilde Y} {\tilde V}^2 \cr 
{\tilde X}&= 3 {\tilde R}^2+3 {\tilde W}\cr
{\tilde W}&= z {\tilde V}^3\cr
{\tilde Y}&= 3 {\tilde R}\cr}}
corresponding to the recursive constructions depicted in Fig. \white.
Eliminating ${\tilde W},{\tilde X},{\tilde Y}$, we arrive at
\eqn\tilmaster{\eqalign{ 
{\tilde R}&={\tilde V} +9 z {\tilde R} {\tilde V}^2 \cr
{\tilde V}&= \theta+ 3 {\tilde R}^2+3 z {\tilde V}^3 \cr}}
Eqns. \tilmaster\ are equivalent to eqns. \master\ upon writing
$R=3 {\tilde V} {\tilde R}$ and $V={\tilde V}$, and may also be 
extracted from the matrix model solution of Ref. \HARD. 
It is quite remarkable that $V={\tilde V}$ has two very different combinatorial interpretations
according to the choice of cutting procedure. Indeed, it may now be viewed as
the generating function for rooted ${\tilde {\rm R}}$-trees starting with an empty vertex
or reduced to a leaf. It can be shown
that this property actually results from the suitable application of the two
different cutting procedures on the {\it same} set of two-leg diagrams with one occupied and one
empty leg, {\it non necessarily lying in the same face}.

Finally repeating the steps of Sects. 3 and 4 above, we find
\eqn\gamttwo{ \gbb  = {\tilde R} - 
{{\tilde R}^3+6 z {\tilde R} V^3\over \theta}  }
and
\eqn\gamtfour{\gbbbb=zV^3+2{\tilde R}^2 -
3{z^2 V^6+{\tilde R}^4+7z {\tilde R}^2V^3 \over \theta} }
in agreement with eqn. \relatgb. 

The existence of two different cutting procedures is a manifestation of the duality 
induced by the scalar product of the underlying two-matrix model
(see Appendix A for more details).

\newsec{Conclusion}

In this paper, we have shown how to recover the matrix model results of Ref. \HARD\ in a purely
combinatorial way, by use of a bijection between configurations of hard particles on tetravalent
planar graphs and suitable trees. 

As a first concluding remark, we note that our
construction is very similar to that used in Ref. \CENSUS, 
where arbitrary rooted planar maps were enumerated. 
Indeed, comparing respectively
the one- and two-leg diagrams (respectively generated by $\Gamma_1$ and $\Gamma_2$) 
of Ref. \CENSUS\ to the two- and four-leg diagrams of the present
paper, we first note that the unrooted trees in bijection with diagrams contributing to $\gnn$
have a simple notion of conjugacy with all their leaves admissible as was the case
for diagrams contributing to $\Gamma_1$. 
On the other hand, the notion of conjugacy for diagrams contributing to 
$\gnnnn$ requires the introduction of the notion of the core of the corresponding trees,
as was also the case when dealing with diagrams contributing to $\Gamma_2$. Moreover,
the various steps used to enumerate the diagrams are exactly parallel in both cases.
In particular, the notion of charge introduced in Ref. \CENSUS\ is a crucial ingredient
for characterizing the balance between buds and leaves within the trees.

The hard-particle model studied here is a special instance of
the more general class of enumeration problems solvable by 
a two-matrix model. The techniques introduced here are easily transposed
to this larger class of problems. We discuss this generalization in
Appendix A where we make the connection between the two-matrix model
formalism and the (suitably generalized) quantities used throughout this paper.

As a last remark, the other models solved in Ref. \HARD\ by matrix techniques, namely 
hard particles on {\it bicolourable} graphs or multicritical versions thereof
(with weaker exclusion rules) should be also amenable to purely combinatorial interpretations
as well.

{\bf Note:} While completing the writing of this paper, we noticed the appearance of a paper
\CONC\ with a similar combinatorial approach to the solutions of the hard-particle
and Ising model on tetravalent planar graphs and with an important overlap with 
our present work.
   
\appendix{A}{Relation between the two-matrix model formalism and the combinatorial approach}
Let us first comment on the relation between the planar solution of the
two-matrix model for hard particles (Sect. 2.2 of Ref. \HARD) and 
the approach of the present paper.
In Ref. \HARD, the solution goes through the introduction of
a family of bi-orthogonal polynomials (eqn. (2.3) of Ref. \HARD ) whose 
norms determine
the all genus partition function through eqn. (2.4) of Ref \HARD. 
These norms may be
computed via algebraic properties of bi-orthogonal polynomials, introducing
in particular the operators $Q_1$ and $Q_2$ describing multiplications 
by eigenvalues. These have finite expansions (eqns. (2.10) of Ref. \HARD) in terms 
of the ``shift" operators acting on the polynomials, and obey 
the ``master equations" (eqns. (2.7) of Ref \HARD) which follow from the scalar product at hand.
The planar limit corresponds to the asymptotics at large polynomial
degree. In this limit, and upon suitable rescalings, the operators 
$Q_1$ and $Q_2^\dagger$ read explicitly
\eqn\qoneto{\eqalign{ Q_1&= \sigma + R \sigma^{-1}+ W \sigma^{-3} \cr
Q_2^\dagger &= V \sigma^{-1} +{{\tilde R}\over V} \sigma + {{\tilde W}\over V^3}\sigma^3\cr}}
where $\sigma$ is a (commuting) dummy variable inherited from the shift 
operator, allowing to keep
track of the degree of the various operators at hand, and $Q_2^\dagger$
is the adjoint of $Q_2$.

In our language, 
the coefficients of $Q_1$ and $Q_2^\dagger$ of the various powers 
of $\sigma$ become
generating functions for charged trees, the charge being directly given by
the corresponding power of $\sigma^{-1}$. 
The operators $Q_1$ and $Q_2^\dagger$ are related algebraically, thus implying
graded relations between their coefficients (eqns. (2.13) of Ref. \HARD\ 
with $S\to W$ and ${\tilde S}\to {\tilde W}$). 
In the combinatorial language, this means
that the trees obey recursive relations in which {\it the charge is conserved}. 

The above degree properties are generic in matrix models solvable 
by orthogonal polynomial techniques. Let us discuss in particular the case of
two-matrix models with potential
\eqn\potomat{\eqalign{ {\cal U}(A,B)&= AB - U(A) -{\tilde U}(B) \cr   
U(A)&= \sum_{i=1}^k g_{i} {A^{2i} \over 2i} \cr
{\tilde U}(B)&=  \sum_{i=1}^k {\tilde g}_{i} {B^{2i} \over 2i} \cr}}
describing in the planar limit bipartite graphs with arbitrary even
vertex valencies and a weight $g_i$ (resp. ${\tilde g}_i$) per black (resp. white)
$2i$-valent vertex, $i=1,2,...,k$.
In this case, the operators $Q_1$ and $Q_2^\dagger$ read in the planar limit
\eqn\plaqoneto{\eqalign{
Q_1&= \sigma + \sum_{i=1}^k {R_i \over \sigma^{2i-1}} \cr
Q_2^\dagger&= {V\over \sigma} +\sum_{i=1}^k {{\tilde R}_i \over V^{2i-1}} \sigma^{2i-1}\cr}}
In the combinatorial language, using again the cutting procedure of Sects. 2 and 3 
(with the convention white=empty and black=occupied) 
for the two-leg diagrams with two black endpoints, we generate bipartite unrooted
R$_1$-trees of charge $+2$, with leaves (resp. buds) only connected
to white (resp. black) vertices,  
whose leaves are all admissible, and whose rooted versions have charge $+1$,
start with a white vertex, and are generated by $R_1$. 
The subtrees of rooted R$_1$-trees 
are rooted trees that either start with a white vertex and have charges $1,3,...,2k-1$,
or with a black vertex and have charges $1,-1,-3,...,-(2k-3)$. The former are generated
by $R_1,R_2,...,R_k$ respectively, while the latter are generated by $V-\theta$, $X_1, X_2,...,X_{k-1}$,
where $X_i= {\tilde R}_i/V^{2i-1}$ and $\theta$ is a weight per leaf. 
These functions are nothing but the coefficients
of $Q_1$ (for white starting vertex) and $Q_2^\dagger$ (for black starting vertex). Note that
the leading coefficients of $Q_1$ and $Q_2^\dagger$ are trivial, with in 
particular $X_k=g_k$ standing for the only tree of charge $-(2k-1)$ starting
with a black vertex, made of $2k-1$ descending buds. 
All these trees obey recursive relations which may immediately be obtained  
by listing all possible vertex descendents (chosen among buds, leaves and
rooted trees in the above list) allowed by charge conservation and 
bipartite character. In the matrix language, these relations read  
\eqn\syst{ \eqalign{
\theta \delta_{m,-1} &= \left( Q_2^\dagger  -U'(Q_1)\right)\bigg\vert_{m}\cr 
\theta \delta_{m,-1} &=  \left( Q_1^\dagger  -{\tilde U}'(Q_2)\right)
\bigg\vert_{m}\cr}}
for $m\geq -1$.
In the above equations, 
the notation $\vert_k$ stands for the coefficient of $\sigma^k$
in the corresponding expression. 

It is important to notice that the charge
restrictions on the above rooted trees translate into the characterizing
property for
unrooted R$_1$-trees that cutting any edge separates the tree into 
two pieces of charges $(1,1)$, $(-1,3)$,..., or $(-(2k-3),2k-1)$, 
the piece of positive charge starting with a white vertex. 

We may then obtain the generating functions for $2i$-leg  
diagrams. For instance,
the two- four- and six-leg diagrams with all-black or all-white legs,
with weights $g_i \theta^{i-1}$ (resp. ${\tilde g}_i\theta^{i-1}$) per
$2i$-valent black (resp. white) vertex and an extra weight of 
$\sqrt{\theta}$ per leg, are respectively generated by:
\eqn\tofourgen{\eqalign{
\gnn &= R_1 - {U'(Q_1)\vert_{-3} \over \theta} \cr 
\gnnnn &= R_2+2 R_1^2 -{U'(Q_1)\vert_{-5} +3 R_1 U'(Q_1)\vert_{-3}\over \theta} \cr
\gnnnnnn &= R_3+6 R_1R_2+5 R_1^3 -{U'(Q_1)\vert_{-7}+5 R_1 U'(Q_1)\vert_{-5}
+(3R_2+9R_1^2) U'(Q_1)\vert_{-3}\over \theta} \cr  
\gbb &= {\tilde R}_1 - {{\tilde U}'(Q_1)\vert_{-3} \over \theta} \cr 
\gbbbb &= {\tilde R}_2+2 {\tilde R}_1^2 -{{\tilde U}'(Q_1)\vert_{-5} +3 {\tilde R}_1 
{\tilde U}'(Q_1)\vert_{-3}\over \theta} \cr
\gbbbbbb &= {\tilde R}_3+6 {\tilde R}_1{\tilde R}_2+5 {\tilde R}_1^3
-{{\tilde U}'(Q_1)\vert_{-7} +5 {\tilde R}_1 
{\tilde U}'(Q_1)\vert_{-5}+(3{\tilde R}_2+9{\tilde R}_1^2) 
{\tilde U}'(Q_1)\vert_{-3}\over \theta} \cr}}
The collection of generating functions 
$\displaystyle{\Gamma_{2i}^{\hbox{$\scriptstyle \n \cdots \n$}\kern -15.2pt
\raise -4pt \hbox{$\scriptstyle \n \cdots \n$}}}$ or that of
$\displaystyle{\Gamma_{2i}^{\hbox{$\scriptstyle \b \cdots \b$}\kern -15.2pt
\raise -4pt \hbox{$\scriptstyle \b \cdots \b$}}}$ for $i=1,2,\cdots,k$
is required to get the generating function 
for the corresponding rooted maps (with no legs), which reads
\eqn\justiE{\eqalign{E& ={\gnb -\theta \over \theta}\cr
\gnb& = \theta + \sum_{i=1}^k g_i \Gamma_{2i}^{\hbox{$\scriptstyle 
\n \cdots \n$}\kern -15.2pt
\raise -4pt \hbox{$\scriptstyle \n \cdots \n$}}  =
\theta + \sum_{i=1}^k {\tilde g}_i \Gamma_{2i}^{\hbox{$\scriptstyle 
\b \cdots \b$}\kern -15.2pt
\raise -4pt \hbox{$\scriptstyle \b \cdots \b$}}  \cr}}
where the equations for diagrams with all-white legs result from
an alternative cutting procedure using
\item{(ii'')} the next visited vertex is black 
\par
\noindent Note finally an obvious black/white duality in which $A\leftrightarrow B$,
$Q_1\leftrightarrow Q_2$, $g_i \leftrightarrow {\tilde g}_i$, 
and $R_i \leftrightarrow {\tilde R}_i$.

\fig{Correspondence between the generating functions in the general
bipartite case and in the tetravalent hard-particle case.
Note that wiping out the two-valent black vertices relaxes the bipartite
constraint into the hard-particle one.}{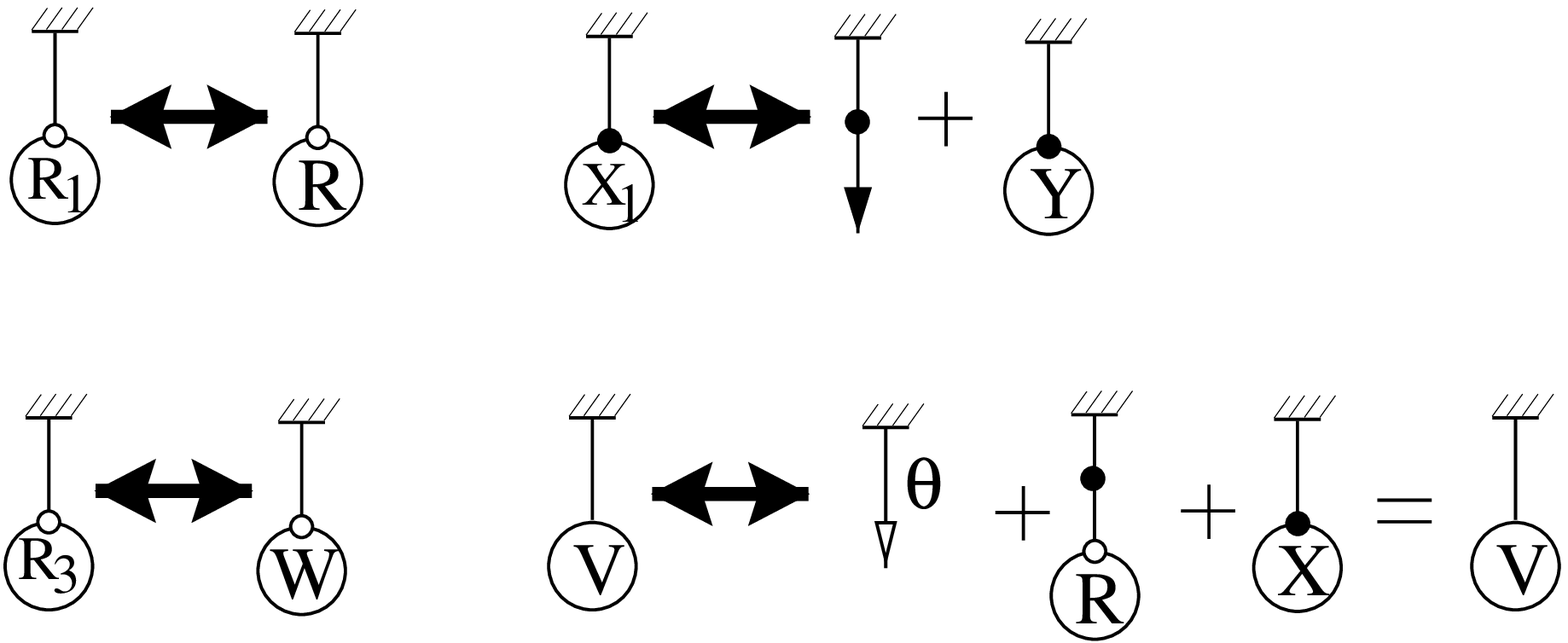}{10.cm}
\figlabel\corresp

The case of hard particles on tetravalent graphs is recovered in this language
by setting $k=2$, $g_1=1$, ${\tilde g}_1=0$ $g_2 =\theta z$, ${\tilde g}_2=\theta$,
and by wiping out the bivalent black vertices so as to build direct edges
between white tetravalent vertices (in this case the condition (ii'')
becomes (ii') of Sect. 5). We have in this case the correspondence   
$R_1\to R$, $R_3\to W$, $X_1\to 1+Y$ and $V\to V$ displayed in Fig. \corresp.

\listrefs
\end